\documentclass{emulateapj}
\pdfoutput=1
\usepackage{longtable}
\usepackage{amsmath}
\usepackage{amssymb}
\usepackage{epsfig}
\usepackage{natbib}

\renewcommand\plotone[1]{%
\typeout{Plotone included the file #1}
\centering
\leavevmode
\includegraphics[clip=true,width=7.5cm]{#1}%
}%
\def\iso#1#2{\mbox{${}^{#2}{\rm #1}$}}

\newcommand{\beq}{\begin{equation}}
\newcommand{\eeq}{\end{equation}}
\newcommand{\beqar}{\begin{eqnarray}}
\newcommand{\eeqar}{\end{eqnarray}}
\def\fun#1#2{\lower3.6pt\vbox{\baselineskip0pt\lineskip.9pt
  \ialign{$\mathsurround=0pt#1\hfil##\hfil$\crcr#2\crcr\sim\crcr}}}%less than approximately and greater than approximately

\begin{document}

\title{Dependence of X-Ray Burst Models on Nuclear Masses}

\author{H. Schatz\altaffilmark{1,2,3} and
W.-J. Ong \altaffilmark{1,2,3}
}

\altaffiltext{1}{National Superconducting Cyclotron Laboratory,
Michigan State University, East Lansing, MI 48824}

\altaffiltext{2}{Joint Institute for Nuclear Astrophysics (JINA),
http://www.jinaweb.org}

\altaffiltext{3}{Department of Physics and Astronomy, Michigan
State University, East Lansing, MI 48824}

\begin{abstract}
X-ray burst model predictions of light curves and final composition of the nuclear ashes are affected by uncertain nuclear masses. However, not all of these masses are determined experimentally with sufficient accuracy. Here we identify remaining nuclear mass uncertainties in X-ray burst models using a one zone model that takes into account the changes in temperature and density evolution caused by changes in the nuclear physics. Two types of bursts are investigated - a typical mixed H/He burst with a limited rp-process and an extreme mixed H/He burst with an extended rp-process. When allowing for a 3$\sigma$ variation only three remaining nuclear mass uncertainties affect the light curve predictions of a typical H/He burst ($^{27}$P, $^{61}$Ga, and $^{65}$As), and only three additional masses affect the composition strongly ($^{80}$Zr, $^{81}$Zr, and $^{82}$Nb). A larger number of mass uncertainties remains to be addressed for the extreme H/He burst with the most important being $^{58}$Zn, $^{61}$Ga, $^{62}$Ge, $^{65}$As, $^{66}$Se, $^{78}$Y, $^{79}$Y, $^{79}$Zr, $^{80}$Zr, $^{81}$Zr, $^{82}$Zr, $^{82}$Nb, $^{83}$Nb, $^{86}$Tc, $^{91}$Rh, $^{95}$Ag, $^{98}$Cd, $^{99}$In, $^{100}$In, and $^{101}$In.  The smallest mass uncertainty that still impacts composition significantly when varied by 3$\sigma$ is $^{85}$Mo with 16 keV uncertainty.  For one of the identified masses, $^{27}$P, we use the isobaric mass multiplet equation (IMME) to improve the mass uncertainty, obtaining an atomic mass excess of -716(7) keV.  The results provide a roadmap for future experiments at advanced rare isotope beam facilities, where all the identified nuclides are expected to be within reach for precision mass measurements.
\end{abstract}

\maketitle

\section{Introduction}
\label{sect:introduction}
Type I X-ray bursts are frequently observed thermonuclear explosions on the surface of neutron stars that accrete matter from a nearby companion star \citep{Schatz2006a,Strohmayer2006,Lewin93,Parikh2013}. The bursts are powered by nuclear reaction sequences that transform accreted hydrogen and helium into heavier elements via the 3$\alpha$-reaction, which burns helium into carbon, the $\alpha$p-process, a sequence of proton captures and ($\alpha$,p) reactions, and the rapid proton capture process (rp-process), a sequence of proton captures and $\beta^+$-decays \citep{Wallace1981,
  VanWormer1994,Schatz1998,Schatz2001,Fisker2008,Woosley2004a,Jose2010}. Nuclear data on neutron deficient rare isotopes are needed to predict burst light curves that can then be compared with observations to constrain system parameters and neutron star properties \citep{Heger2007,Galloway2004,Zamfir2012}. Nuclear data are also needed to calculate the composition of the burst ashes to predict possible composition specific spectral signatures \citep{Weinberg2006,NUSTAR2015,Kajava2016}, and to predict the composition of the neutron star crust, which in turn influences heat transport properties as well as strength and distribution of various deep nuclear heating and cooling processes \citep{Haensel2008,Gupta2007, Schatz2014}. This relates to observations of crustal cooling during quiescence in transiently accreting systems, which can provide unique insights into the stellar interior of neutron stars \citep{Brown2009,Horowitz2015}. 

Nuclear masses play an important role in X-ray burst models as they define the location of the proton drip line and therefore the path of the rp-process \citep{Schatz1998,Schatz2006}. Motivated by the data needs of X-ray burst models, a large number of mass measurements on very neutron deficient isotopes  have been carried out by taking advantage of new radioactive beam production capabilities and advances in experimental techniques such as Penning traps \citep{Clark2004,Rodriguez2004,Schury2007,Clark2007,Weber2008,Savory2009,Elomaa2009,Haettner2011,Fallis2011,Kankainen2012} and storage rings \citep{Stadlmann2004,Tu2011a,Yan2013}. These techniques provide mass data with the required accuracy of better than 10-100~keV or about 1:10$^6$. Measurements have also reached beyond the proton drip line using $\beta$-delayed proton spectroscopy \citep{DelSanto2014}  and proton breakup \citep{Rogers2011}. In addition, improvements in nuclear theory enable the calculation of the masses of the most exotic rp-process nuclei for which the proton number exceeds the neutron number with an accuracy of about 100~keV \citep{Brown2002}. This approach uses predictions of Coulomb shifts to calculate masses from the measured mass of the less exotic mirror nucleus, where proton and neutron numbers are exchanged. Because of these developments, the need for  global mass models has largely been eliminated, and as a consequence the uncertainties of nuclear masses along the rp-process are now rather well characterized, and correlations among uncertainties are much reduced. This paper takes advantage of the improved knowledge of nuclear masses to quantify the impact of the remaining mass uncertainties on X-ray burst models, and to provide guidance for future measurements that address them. 

Nuclear masses $m$ enter X-ray burst models directly in form of proton capture Q-values $Q_{(p,\gamma)}=m(Z,A)+m_p-m(Z+1,A+1)$ (with nuclei of mass number $A$ and charge number $Z$ and proton mass $m_p$).  $Q_{(p,\gamma)}$ values are used to calculate  the  ($\gamma$,p) photodisintegration rates $\lambda_{(\gamma,p)}$ from 
the (p,$\gamma$) proton capture rates $<\sigma v>_{(p,\gamma)}$ via detailed balance
\begin{equation} \label{eq:detailed}
\lambda_{(\gamma,p)} = \frac{2G_f}{G_i} \left( \frac{\mu kT}{2 \pi \hbar^2} \right)^{3/2} \exp \left( -\frac{Q_{(p,\gamma)}}{kT} \right)
<\sigma v>_{(p,\gamma)}
\end{equation} with partition functions of initial and final nuclei $G_i$ and $G_f$, reduced mass $\mu$, and temperature $T$ \citep{Schatz1998}.
The ratio of the ($\gamma$,p) reaction rate to the reverse  (p,$\gamma$) rate depends exponentially on $Q_{(p,\gamma)}$  and is therefore very sensitive to nuclear masses. The ratio strongly increases as proton captures reach more and more proton rich nuclei with lower Q-values, and eventually becomes large enough to impede the proton capture process, which then has to wait for a slow $\beta^+$ decay to occur before proton captures can resume. Around these so called waiting points (p,$\gamma$)  and ($\gamma$,p) reactions compete with each other. In this situation the net reaction flow depends directly on the ratio of the competing rates and therefore on nuclear masses. 

In principle, nuclear mass uncertainties can also enter X-ray burst models as part of reaction rate uncertainties. Reaction rates are for the most part not measured directly but calculated, and in many cases this requires the input of nuclear masses. These cases include reaction rates predicted using the Hauser-Feshbach statistical approach, which depends on reaction Q-values, and reaction rates calculated from resonance properties, if resonance energies are not determined directly but deduced from excitation energies and the reaction Q-value \citep{Schatz2006,Wrede2010a}. However, this uncertainty is included in previous sensitivity studies, which took into account reaction rate uncertainties due to all nuclear structure ingredients, including masses \citep{Cyburt2016}. This work complements that study and focuses exclusively on the additional direct dependence of X-ray burst models on nuclear masses via the $Q_{(p,\gamma)}$ 
factor of Eq.~\ref{eq:detailed} in situations of competing forward and reverse rates. This effect has not been included in reaction rate sensitivity studies such as \citet{Cyburt2016}, which vary forward and reverse reactions together according to detailed balance (Eq.~\ref{eq:detailed}) and keep Q-values fixed. Our approach neglects correlations between these additional mass uncertainties, and the reaction rate uncertainties. This is justified as during local (p,$\gamma$)-($\gamma$,p) equilibrium, when the mass uncertainty contribution studied in this work is most important, reaction rates and their uncertainties become unimportant and vice versa. 

The importance of nuclear masses for rp-process calculations was shown in \citet{Schatz1998} who performed constant temperature and density calculations with different mass models. A single zone X-ray burst model similar to the one of the present study has been used to demonstrate the strong impact of nuclear mass uncertainties in specific cases, usually in the context of new mass measurements, including $^{68,70}$Se \citep{Savory2009}, $^{105}$Sn and $^{106}$Sb \citep{Elomaa2009}, $^{65}$As \citep{Tu2011a}, $^{69}$Br \citep{DelSanto2014}, and
$^{45}$Cr \citep{Yan2013}. Only one pioneering large scale systematic study of mass uncertainties has been carried out before \citep{Parikh2009}, based on the 2003 Atomic Mass Evaluation (AME2003) \citep{Audi2003}. However, a large number of masses have been measured since that time. In addition the study had some limitations that are overcome in this work: It used the post-processing approach, where the impact of mass uncertainties on the temperature and density evolution is neglected and which therefore cannot determine the impact of mass uncertainties on burst light curves. The study was also limited to varying Q-values of less then 1~MeV. 

\section{Method}

The sensitivity of X-ray burst models to nuclear masses is analyzed using a one-zone model \citep{Schatz2001,Cyburt2016} that has been shown to predict nuclear processes, light curves, and final composition of the burst ashes with sufficient similarity to full 1D models  to be useful to identify important nuclear uncertainties. X-ray bursts in nature show a broad range of characteristics depending on accretion rate, accreted composition, and neutron star properties. As a consequence, nuclear processes can differ significantly. In particular, the amount of hydrogen at ignition can vary, which strongly affects the extent of the rp-process.  We use two different ignition conditions to span this range. Model A is characterized by a large initial hydrogen abundance of 0.66. Such ignition conditions would occur in a system that accretes low metallicity material at a relatively high accretion rate and has been used in previous work to map out the possible extent of an rp-process in X-ray bursts \citep{Schatz2001}. The rp-process in model A reaches all the way to the Sn-Sb-Te cycle. Model B is identical to the model ONEZONE in \citet{Cyburt2016} and has been tailored through comparison with a full 1D model to represent the mixed hydrogen and helium bursts observed in GS 1826-24. In model B the main rp-process ends in the $A=60-64$ range, with a weaker reaction flow reaching into the $A=80$ region. Tab.~\ref{tbl:models} summarizes the pressure at ignition depth $P$, initial hydrogen ($X$) and helium ($Y$) mass fractions, and peak temperature $T_{\rm peak}$ for both models. 

Nuclear reaction rates were taken from JINA reaclib V2.0 \citep{Cyburt2010}. Nuclear Q-values were calculated from atomic masses. Experimental atomic masses were taken from the Atomic Mass Evaluation AME2012 \citep{AME12}. Unknown atomic masses beyond the $N=Z$ line were calculated from experimental masses of mirror nuclei using the Coulomb displacement energies from \citet{Brown2002} and adding in quadrature an additional uncertainty of 100~keV. For the remaining nuclei with unmeasured masses, the mass extrapolations provided by AME2012 were used. Using these Q-values, reverse rates were calculated for a particular mass table using Eq.~\ref{eq:detailed}. For the mass variations, only ($\gamma$,p) reactions were recalculated as these are the only cases where reactions in forward and reverse direction compete, and the uncertainty on the ratio of forward and reverse reaction rates of interest here matters. We do not recalculate theoretically predicted forward rates $<\sigma v>_{(p,\gamma)}$ with the modified masses. 

\begin{deluxetable}{lcc}
  \tablecaption{\label{tbl:models} Model parameters}
\tablewidth{0pt}
\tablehead{
\colhead{Parameter} & \colhead{Model A} & \colhead{Model B} 
}
\startdata
$P$ (dyne/cm$^2$)               &    $6.7\times 10^{22}$ & $1.73\times 10^{22}$   \\
$X$   &     0.66     &   0.51       \\
$Y$   &     0.34     &   0.39       \\
$T_{\rm peak}$  (GK)             &     2.0       &   1.2        \\
$N_{\rm nuc}$       &    330       &   263    
\enddata
\end{deluxetable}

For each X-ray burst model, the set of nuclei for which masses were varied was identified by requiring a net reaction flow integrated over the burst duration either leading to or from the nuclide of at least 10$^{-5}$~mole/g. The net reaction flow is defined as 
\begin{equation}
\int_t \left\{ \left( \frac{dY_{\rm i}}{dt} \right)_{{\rm i} \rightarrow {\rm f}} - \left( \frac{dY_f}{dt} \right)_{{\rm f} \rightarrow {\rm i}} \right\} dt
\end{equation}
where $(dY_{\rm i}/dt)_{{\rm i} \rightarrow {\rm f}}$ is the abundance change for initial nuclide i induced by the particular reaction under consideration that converts the initial nuclide i into final nuclide f. 
For comparison, the integrated net reaction flow through the 3$\alpha$ reaction, which is the major bottleneck for creating seed nuclei and can therefore serve as a useful normalization is of the order of 10$^{-2}$~mole/g.  The number of selected nuclides $N_{\rm nuc}$ is listed in Tab.~\ref{tbl:models}. Because of the explicit calculation of reaction flows, (p,$\gamma$)-($\gamma$,p) equilibria tend to result in erroneously large net reaction flows. This is a well-known problem (see for example Fig. 12 in \citet{Cyburt2016} and the unreasonable large flow from $^{59}$Zn to $^{60}$Ga), but helps here as it ensures that nuclei involved in such equilibrium clusters are not missed even if true net flows are weak. This is important, as the influence of mass uncertainties is greatest for nuclei participating in such equilibria. 

For each of the important nuclei, two burst calculations were carried out, one with the mass increased by 3$\sigma$, and one with the mass decreased by 3$\sigma$. Light curve and final composition were then compared to determine the impact of the respective mass uncertainty. Differences in light curves are quantified by the maximum light curve ratio $r_{\rm LC}$ among all time steps, either $\max_t (L_t^{\rm up}/L_t^{\rm down})$ or  $\max_t (L_t^{\rm down}/L_t^{\rm up})$ depending on which one is larger than 1, with time steps $t$ and luminosity $L^{\rm up}$ and $L^{\rm down}$ for the burst light curves obtained with a mass increase or decrease, respectively. 

 Prior to the comparison a time offset is applied to align the light curve peaks in time. This prevents small changes in ignition time that simply shift the burst light curve in time and would be observationally irrelevant, to appear as large discrepancies. However, small changes in light curve shape around the peak that lead to a small offset correction can still lead to spuriously large $r_{\rm LC}$, especially during the steep burst rise. An example is the $^{36}$Ca mass, which has a small effect on the late rise and the shape of the burst peak resulting in a small $r_{\rm LC}$=1.2. However, the burst shape change results in a small 0.14~s time shift of the burst peak and, once one corrects for the shift, an artificially large $r_{\rm LC}$=3.5 results from the steep early burst rise. These issues are identified by visual inspection.  We divide light curve impacts in two qualitative categories based on the judgement of the authors, following the approach of  \citet{Clayton1974} and \citet{Cyburt2016}. Category 1 are impacts that are likely observationally relevant. We use the general criterion $r_{\rm LC}>$2.5, but make exceptions for smaller  $r_{\rm LC}$ values if the shape of the light curve is significantly changed beyond simple changes in burst duration. Category 2 are impacts that are small, but may become significant if uncertainties were larger. We use a threshold of $r_{\rm LC}>$1.7 , below which we judge changes to be unlikely to be observationally relevant. 
 
The final composition is summed by mass number to focus on composition changes during the burst, and not on changes along isobaric chains that may occur towards the end of the burst model calculation due to long lived decays and continuum electron captures. As a measure of impact $r_{\rm Comp} =  y_{\rm up}/y_{\rm down}$ is used, with $y_{\rm up}$ and $y_{\rm down}$ being the final abundances, summed by mass number,  obtained with a mass increase and decrease, respectively. Only $r_{\rm Comp}$ values where at least one of the final abundances is above $10^{-5}$ are considered. 

\section{Results}
\begin{figure*}
\begin{center}
\includegraphics*[clip=true,width=5.3cm]{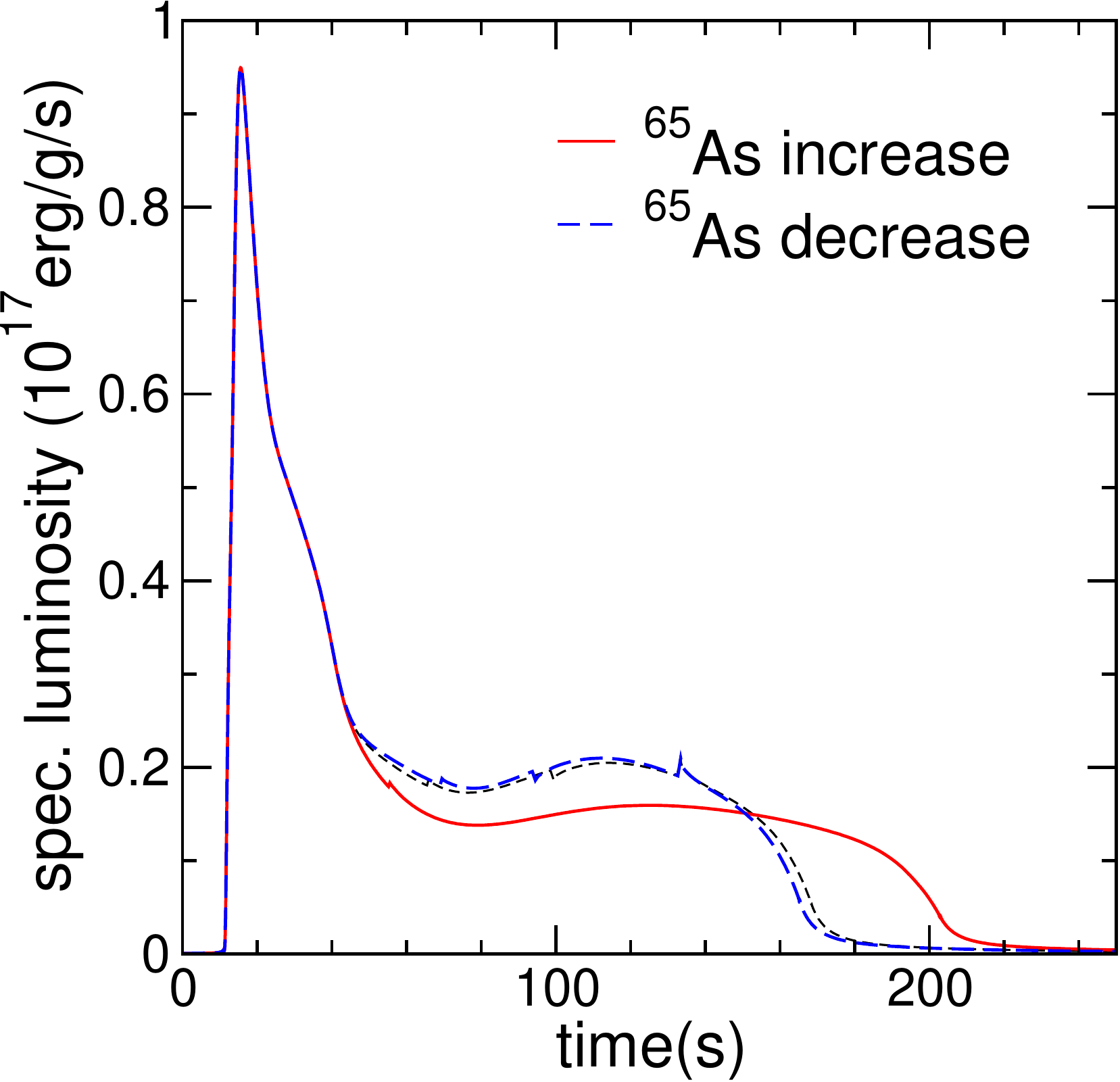}%
\hfill
\includegraphics*[clip=true,width=5.3cm]{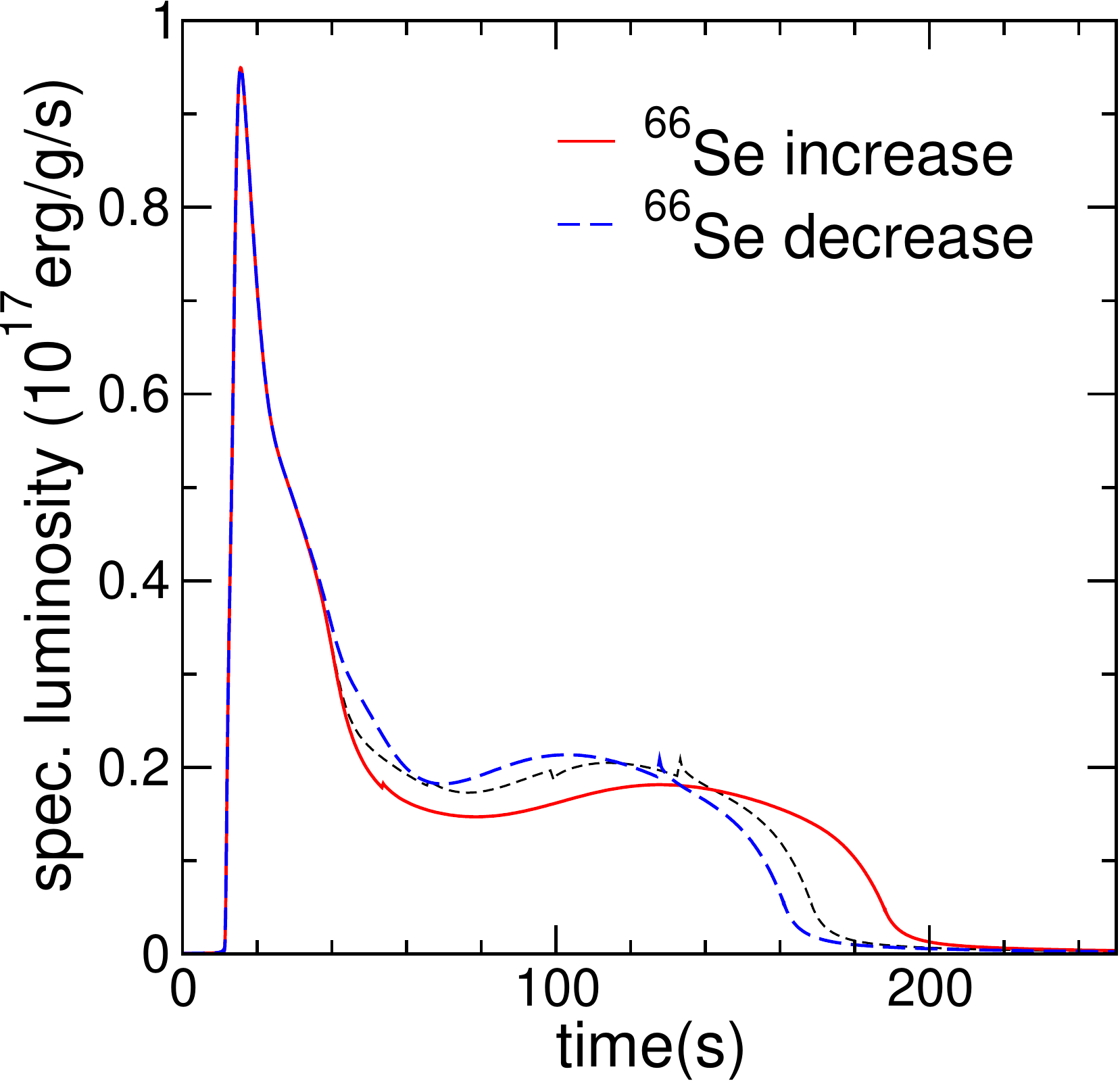}%
\hfill
\includegraphics*[clip=true,width=5.3cm]{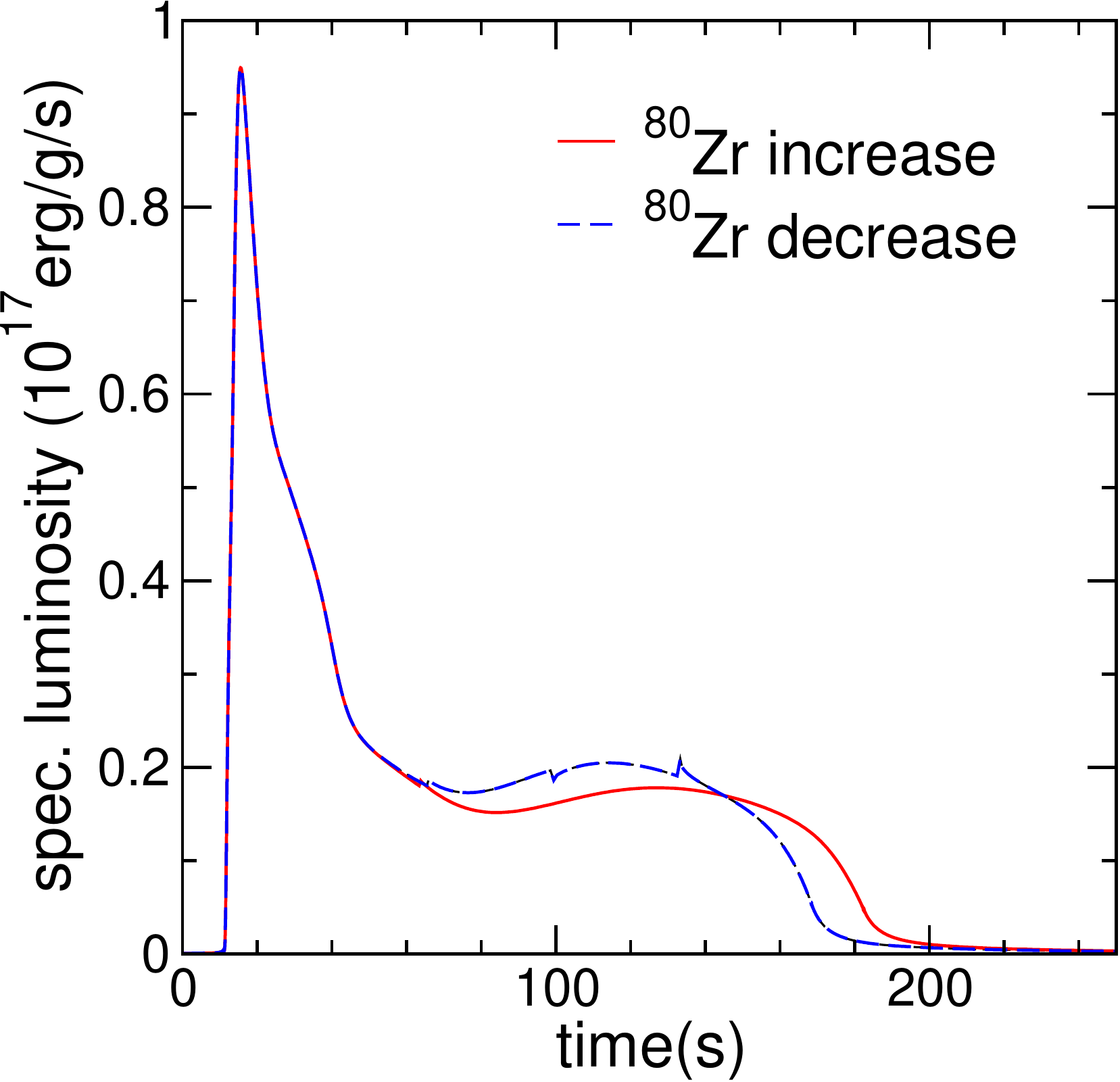}\\
\hfill
\includegraphics*[clip=true,width=5.3cm]{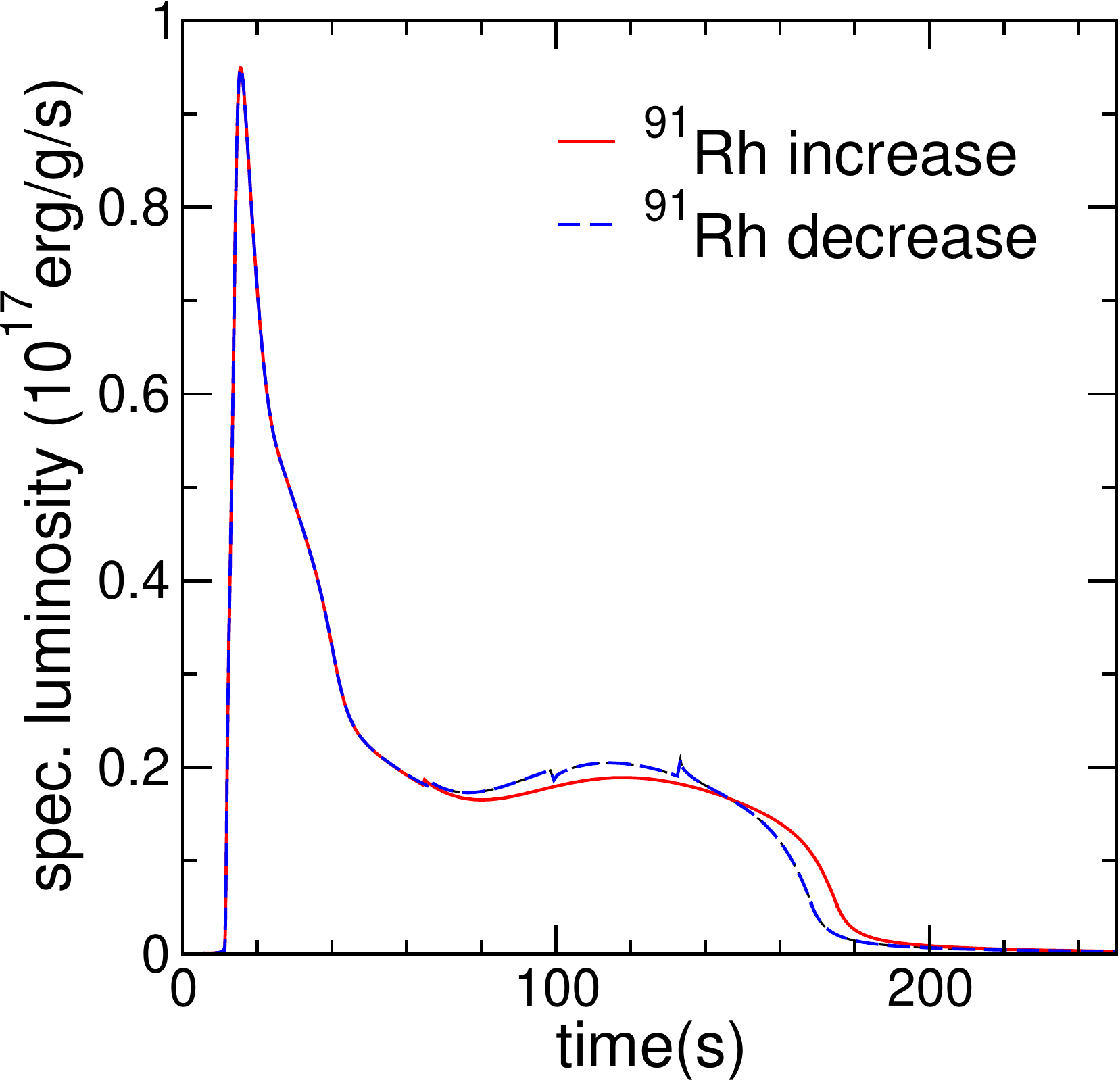}%
\hfill
\includegraphics*[clip=true,width=5.3cm]{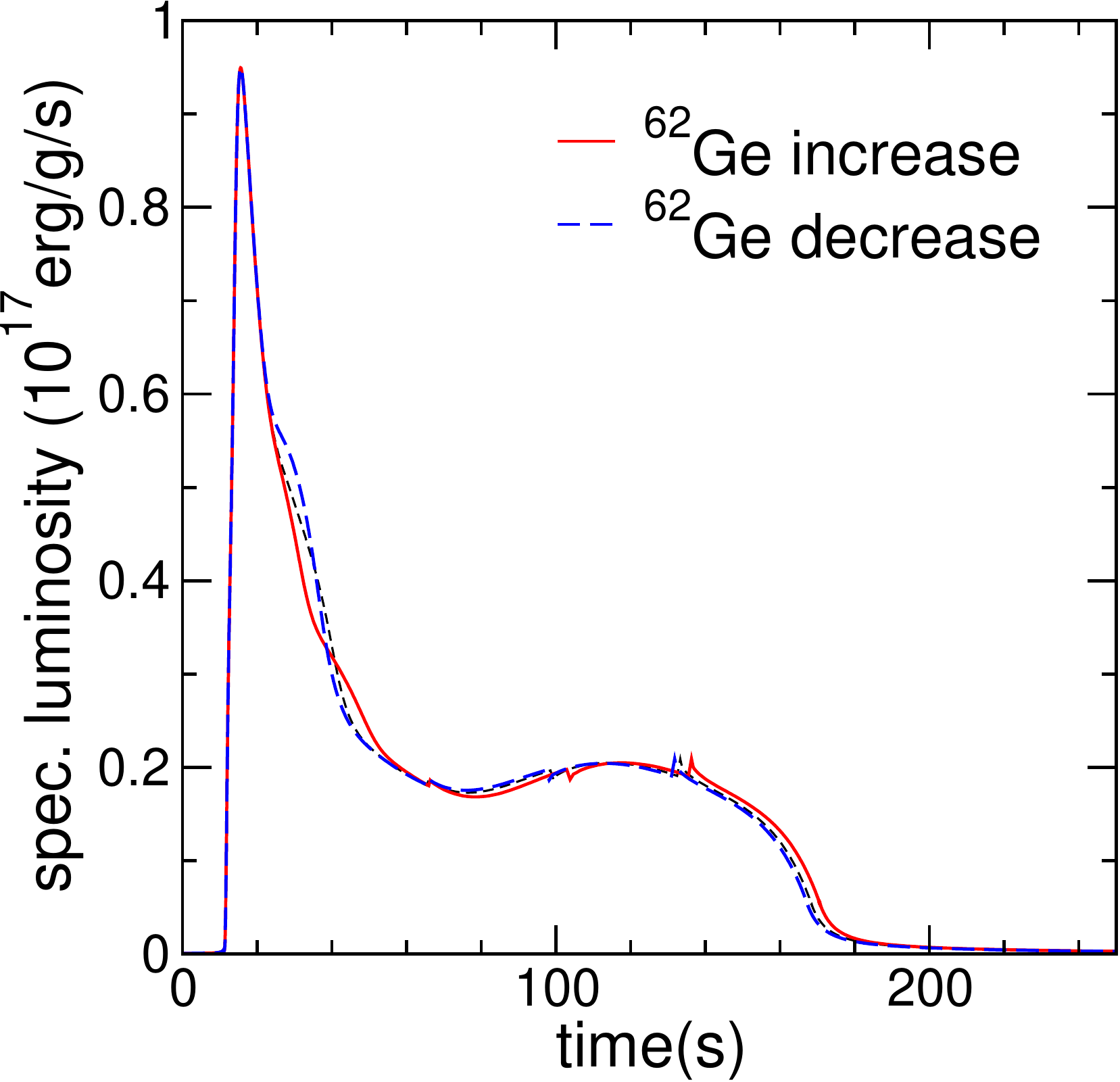}%
\hfill
\includegraphics*[clip=true,width=5.3cm]{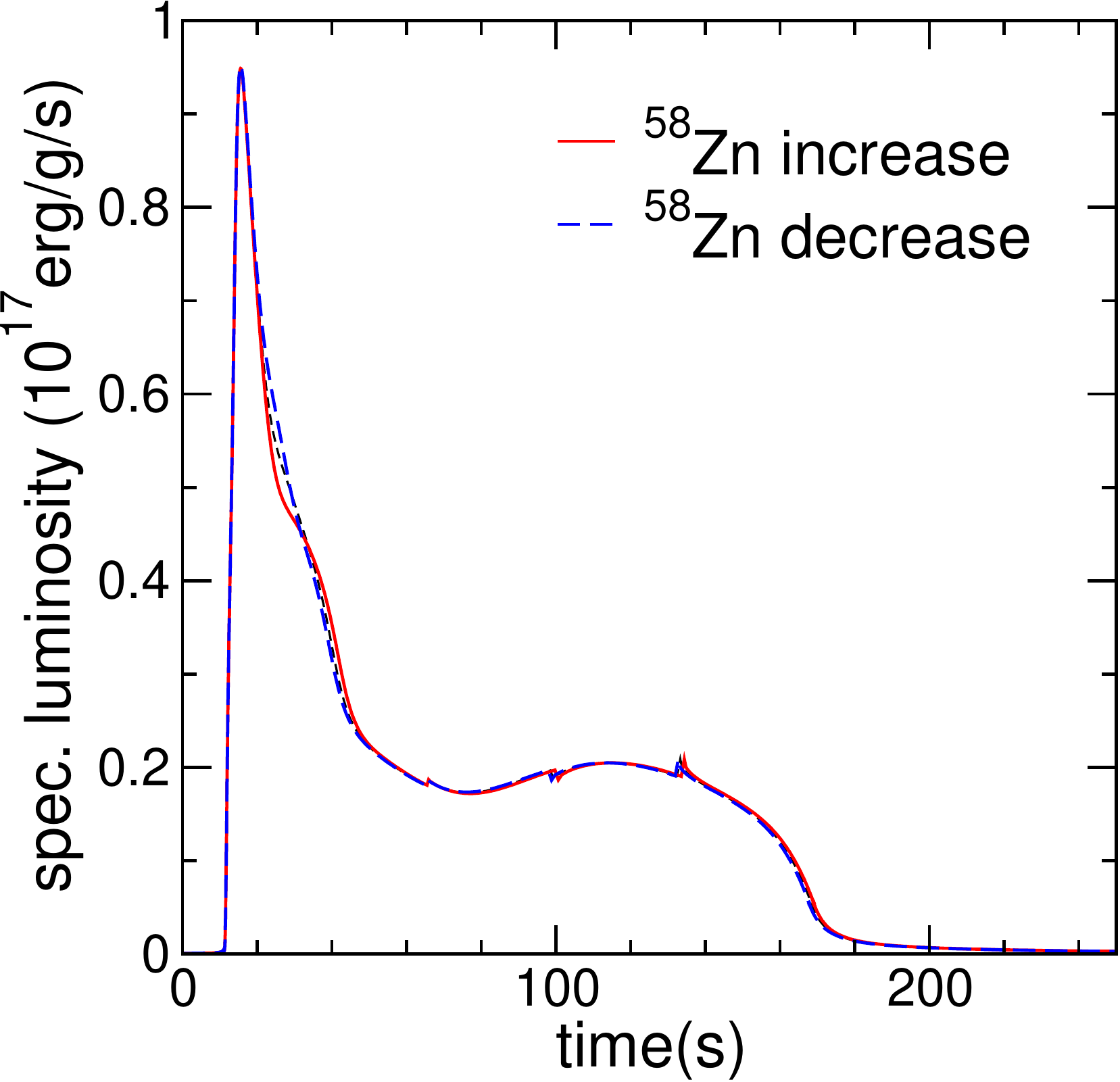}
\end{center}
\caption{\label{FigLC_A} Model A specific luminosities as functions of time for the most significant (Category 1) light curve variations, with baseline (thin, black, dashed), mass increase (red, solid), and mass
decrease (thick, blue, dashed). }
\end{figure*}

\begin{figure*}
\begin{center}
\includegraphics*[clip=true,width=5.3cm]{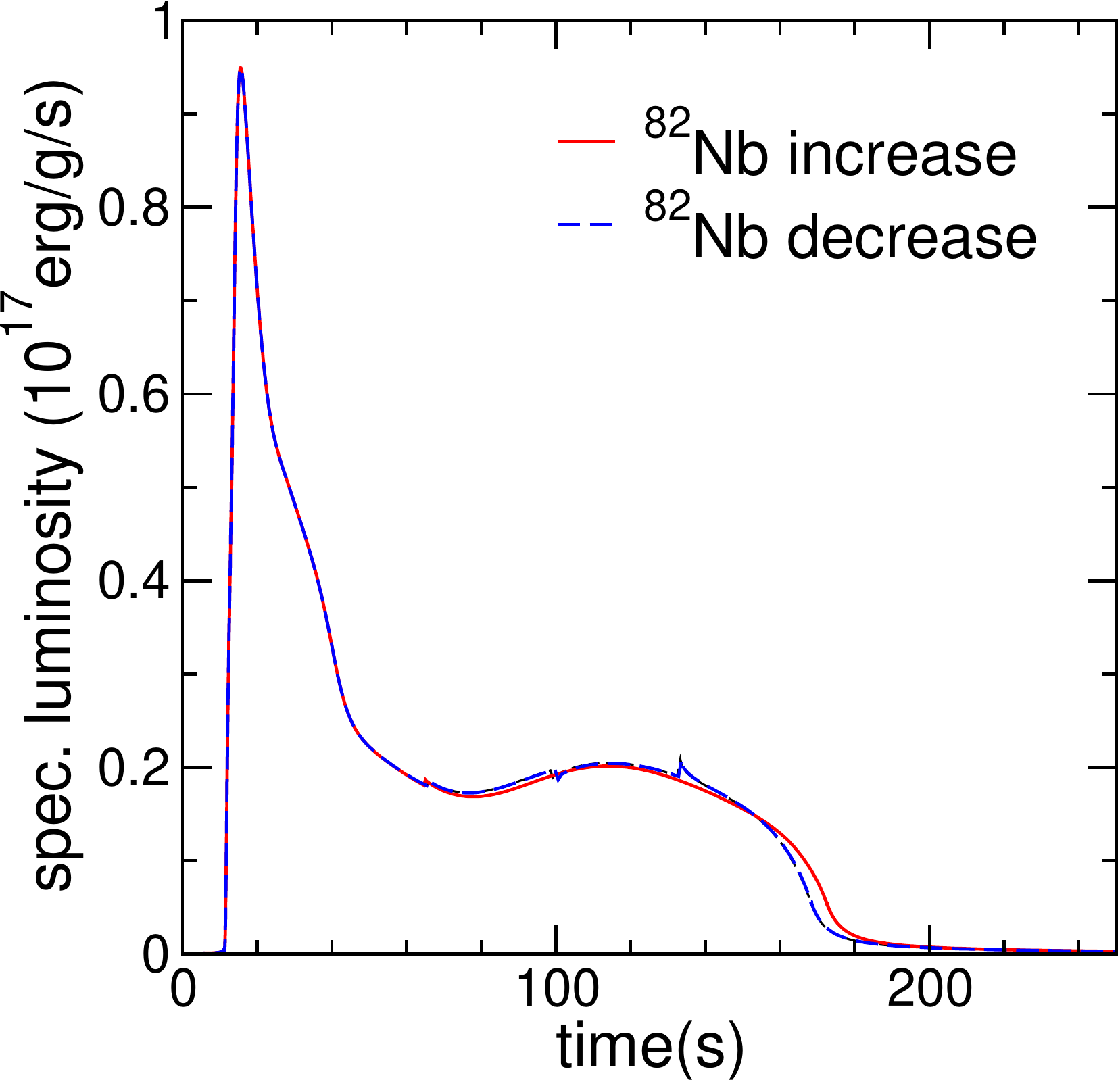}%
\hfill
\includegraphics*[clip=true,width=5.3cm]{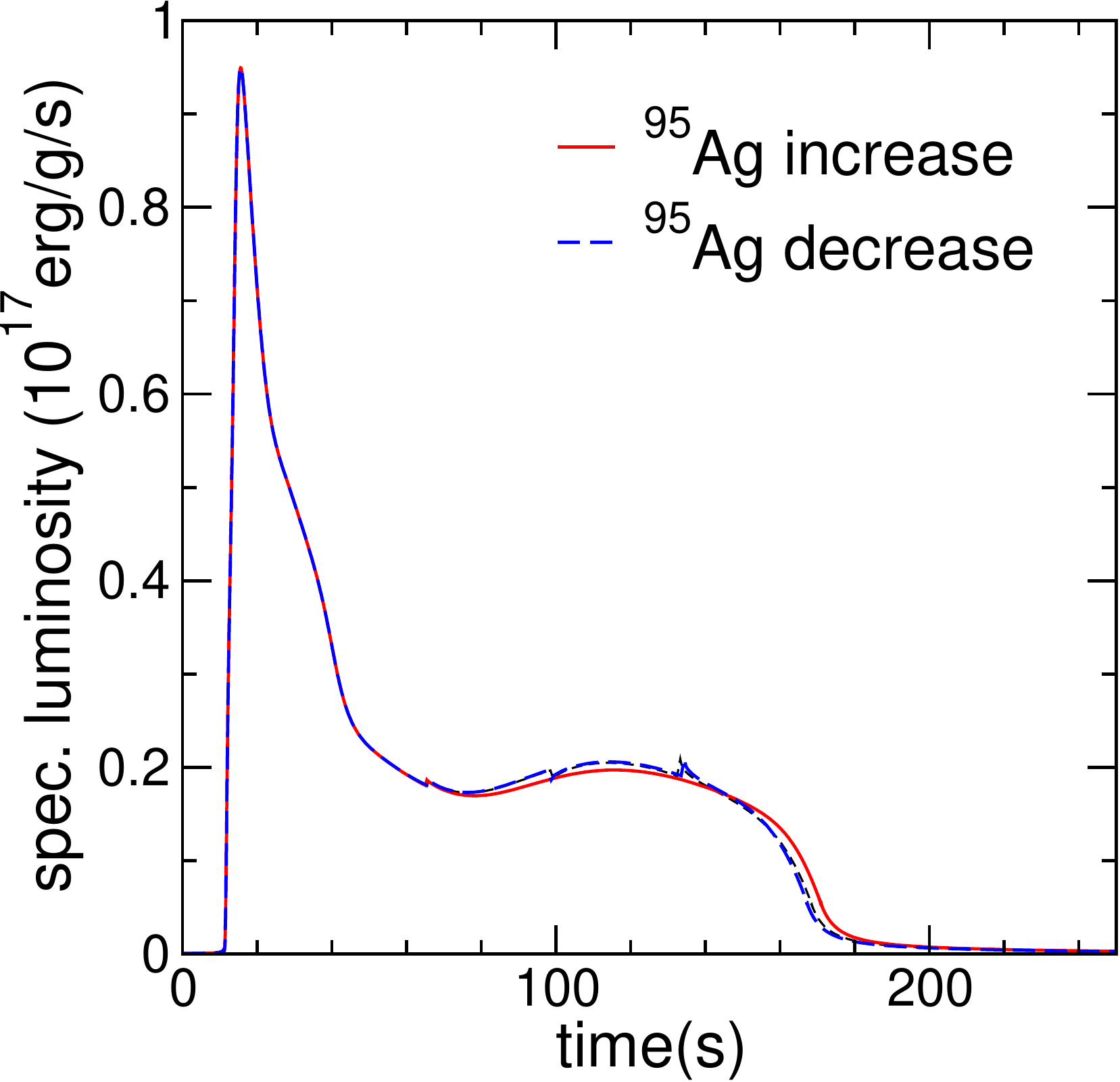}%
\hfill
\includegraphics*[clip=true,width=5.3cm]{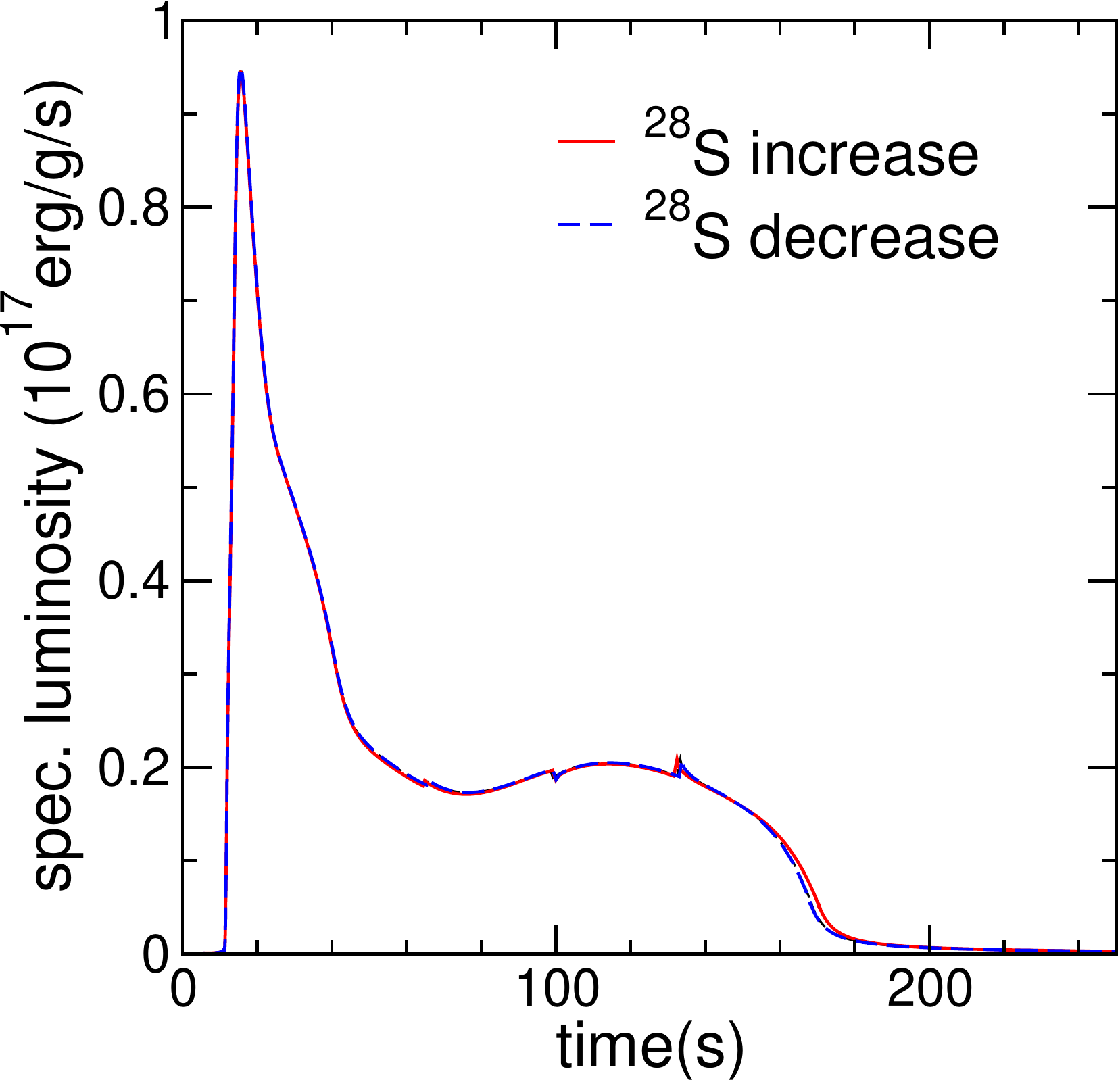}\\
\includegraphics*[clip=true,width=5.3cm]{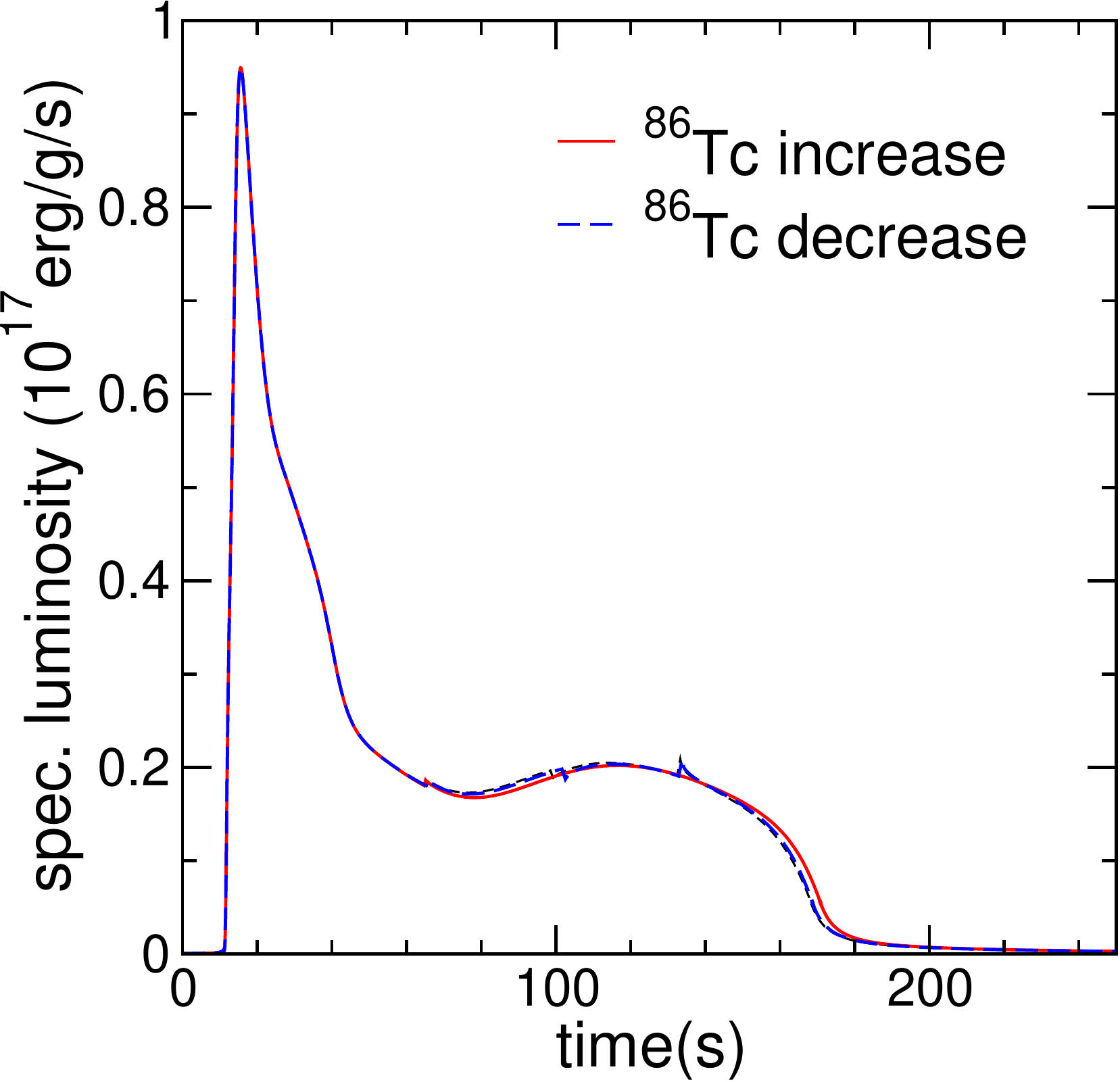}%
\hfill
\includegraphics*[clip=true,width=5.3cm]{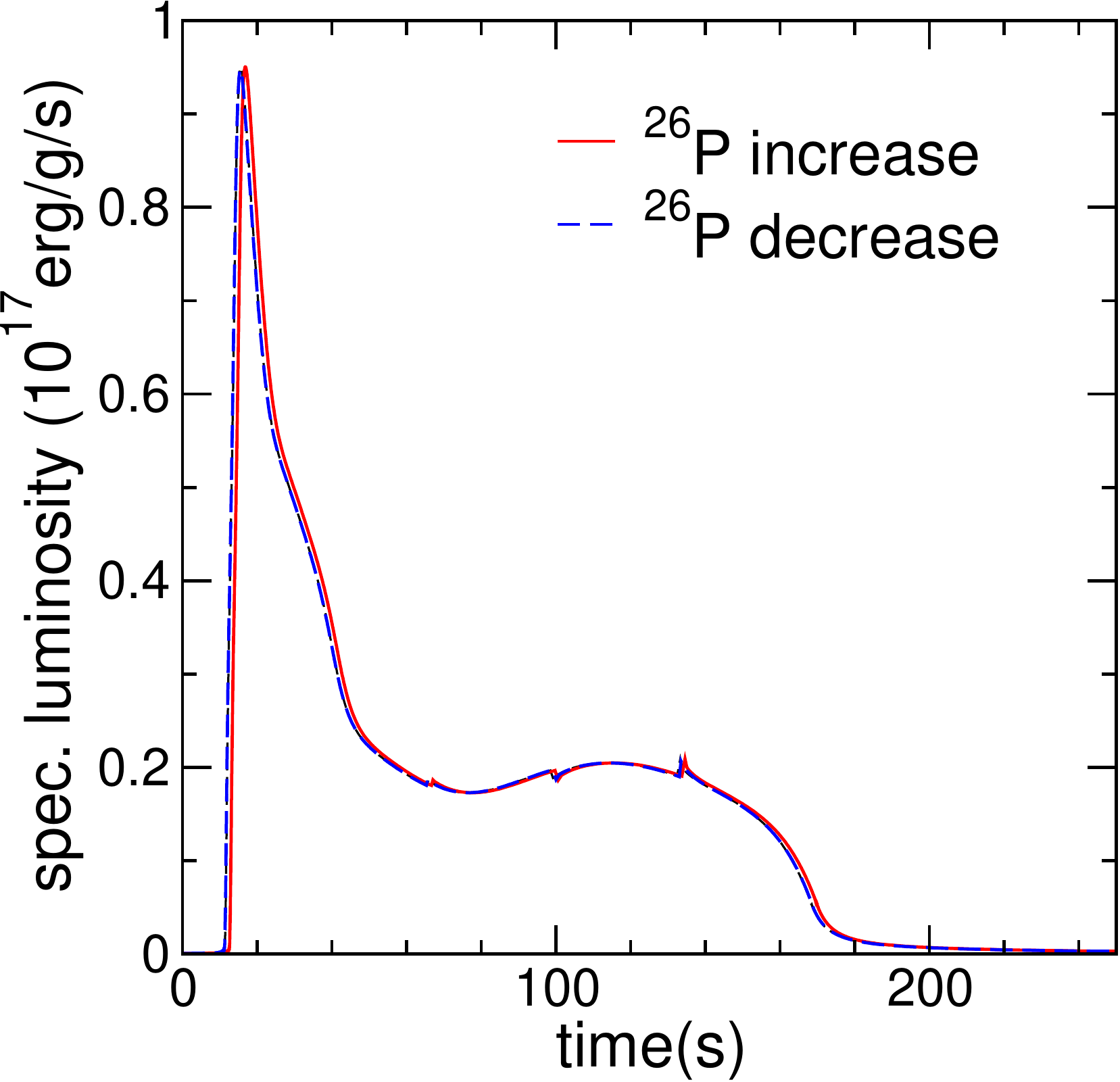}
\end{center}
\caption{\label{FigLC_A2} Model A specific luminosities as functions of time for smaller light curve variations mentioned in the text. See Fig.~\ref{FigLC_A} for details.}
\end{figure*}

The mass uncertainties $\sigma$ impacting the light curve (when masses are varied by 3$\sigma$) for Model A, which has the most extended rp-process and therefore the largest number of relevant mass uncertainties, are listed in Tab.~\ref{tbl:LC} . Only four mass uncertainties have a major (category 1)  effect on the light curve: $^{65}$As, $^{66}$Se, $^{80}$Zr, and $^{91}$Rh (Fig.~\ref{FigLC_A}). Two mass uncertainties, $^{62}$Ge and $^{58}$Zn,  have an interesting effect on the shape of the early light curve cooling, and are therefore also classified as category 1 (Fig.~\ref{FigLC_A}). Two additional mass uncertainties have smaller (category 2) impacts, $^{82}$Nb, and $^{95}$Ag (Fig.~\ref{FigLC_A2}). There are three mass uncertainties not listed in Tab.~\ref{tbl:LC} that produce negligible effects that are barely noticeable in the graphs and that are mentioned for completeness - $^{28}$S ($\sigma=160$~keV, $r_{\rm LC}=$1.5), $^{26}$P ($\sigma=196$~keV, $r_{\rm LC}=$1.3), and $^{86}$Tc ($\sigma=298$~keV, $r_{\rm LC}=$1.5) (Fig.~\ref{FigLC_A2}).

\begin{deluxetable}{crlrrr}
  \tablecaption{\label{tbl:LC} Mass uncertainties that impact the burst light curve
}
\tablewidth{0pt}
\tablehead{
\colhead{Isotope} & \colhead{$\sigma$\tablenotemark{a}} & Source\tablenotemark{b} & \colhead{Model} & \colhead{$r_{\rm LC}$} & \colhead{Cat.} 
}
\startdata
\iso{As}{65}  &    85     &   Exp & A/B & 12.1/2.0  & 1/2 \\
\iso{Se}{66}  &   100    &   CDE & A & 10.3 & 1 \\
\iso{Ga}{61}  &    38     &   Exp & B  & 5.7 & 1 \\
\iso{Zr}{80}   &  1490   &   Exp & A  & 5.0    & 1 \\
\iso{P}{27}    &     26     &   Exp & B  & 4.5 & 1 \\
\iso{Rh}{91}  &    401   &   XTP & A  & 2.9    &  1\\
\iso{Nb}{82}  &    298   &   XTP & A  & 2.0    &  2\\
\iso{Ag}{95}  &    401    &   Exp & A & 1.9    &   2\\
\iso{Ge}{62}  &    100    &  CDE & A &  1.8    &   1\\
\iso{Zn}{58}   &    50     &   Exp & A & 1.7   &   1
\enddata
\tablenotetext{a}{Mass uncertainty in keV. Masses were varied by 3$\sigma$.}
\tablenotetext{b}{Source for mass value: Exp (experimental from AME2012), XTP (extrapolated from AME2012), CDE (Coulomb displacement energy calculation).}
\end{deluxetable}

\begin{figure}
\plotone{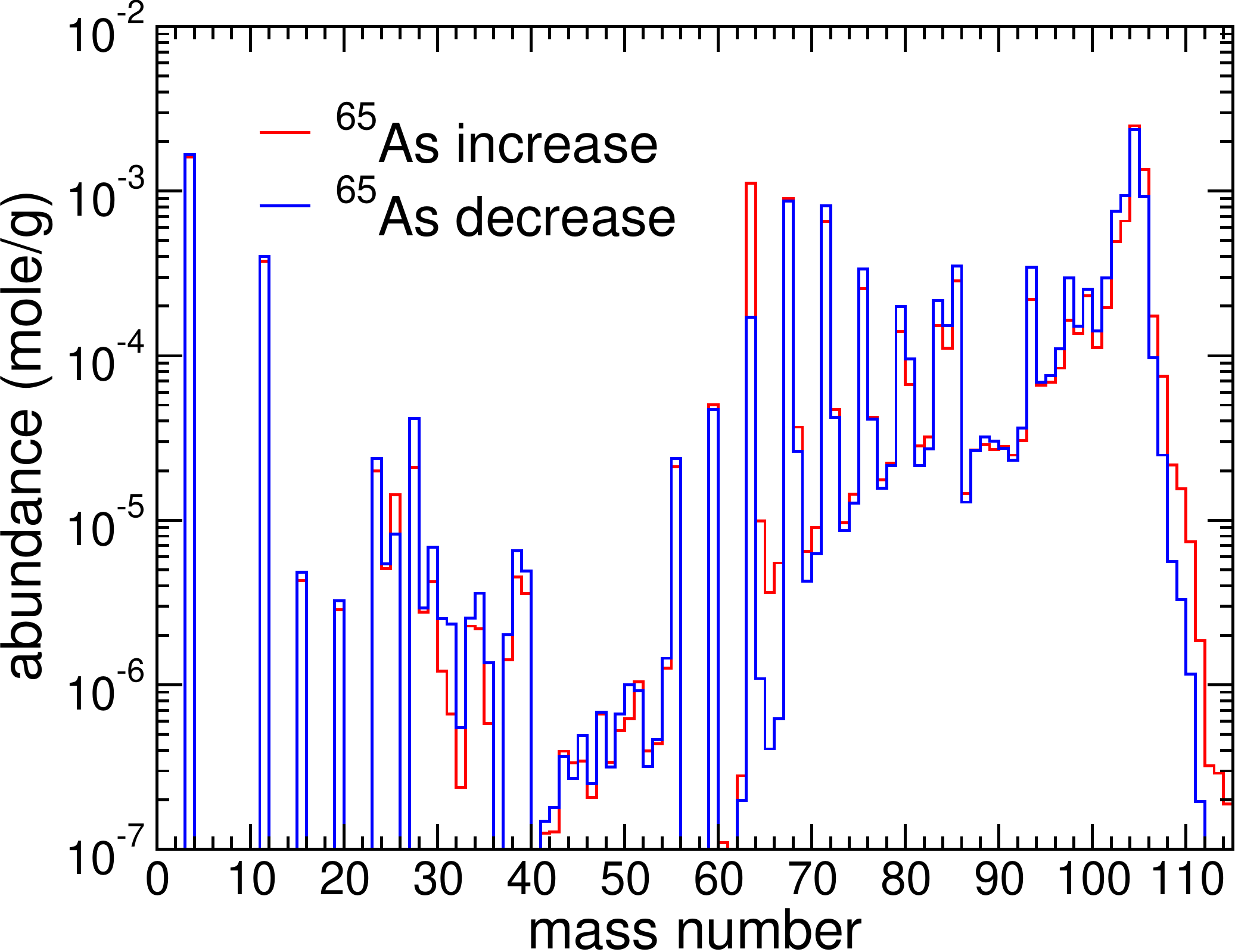}
\caption{\label{FigAB_A} Model A final compositions of the burst ashes summed by mass number for the variation of the $^{65}$As mass.}
\end{figure}

Mass uncertainties that lead to more than a 20\% abundance change in Model A are listed in Tab.~\ref{tbl:AB_A}. Fig.~\ref{FigAB_A} shows an example. 
Mostly, the masses that affect the light curve strongly also affect the composition of the burst ashes significantly. However, there are many additional mass uncertainties listed in Tab.~\ref{tbl:AB_A} that affect composition, but have only a negligible effect on the light curve ($r_{\rm LC}< 1.3$). This is simply a consequence of the fact that delay times at rp-process waiting points, where $\beta^+$ decay occurs, vary widely. Overall energy generation, and therefore the light curve, is controlled by the slowest waiting points. These waiting points control the overall reaction flow and therefore also tend to affect composition broadly. However, on top of this global effect, the abundance produced in an individual mass chain is strongly controlled by the local waiting point at that mass number. 

\begin{figure*}
\begin{center}
\includegraphics*[clip=true,width=5.3cm]{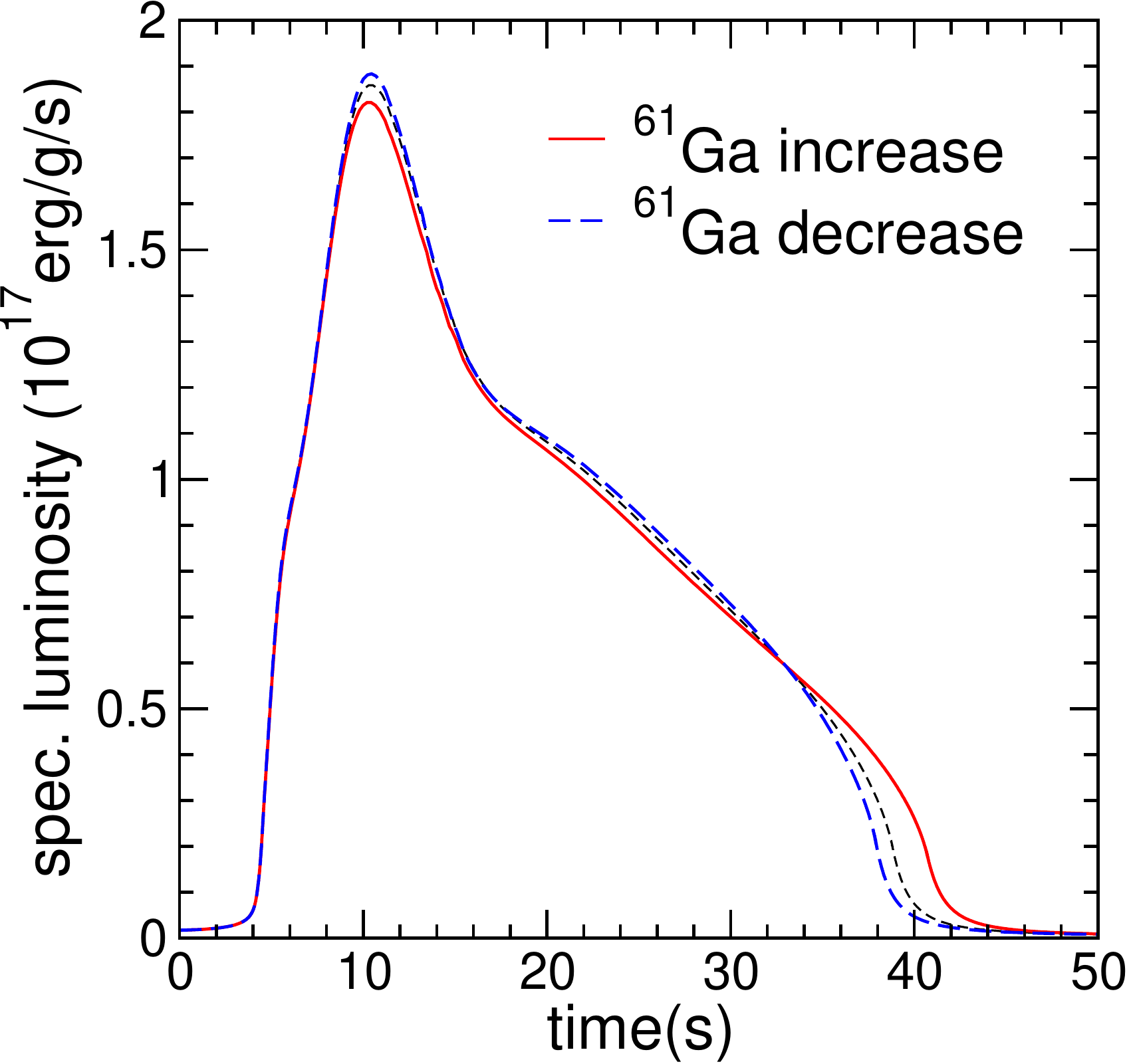}%
\hfill
\includegraphics*[clip=true,width=5.3cm]{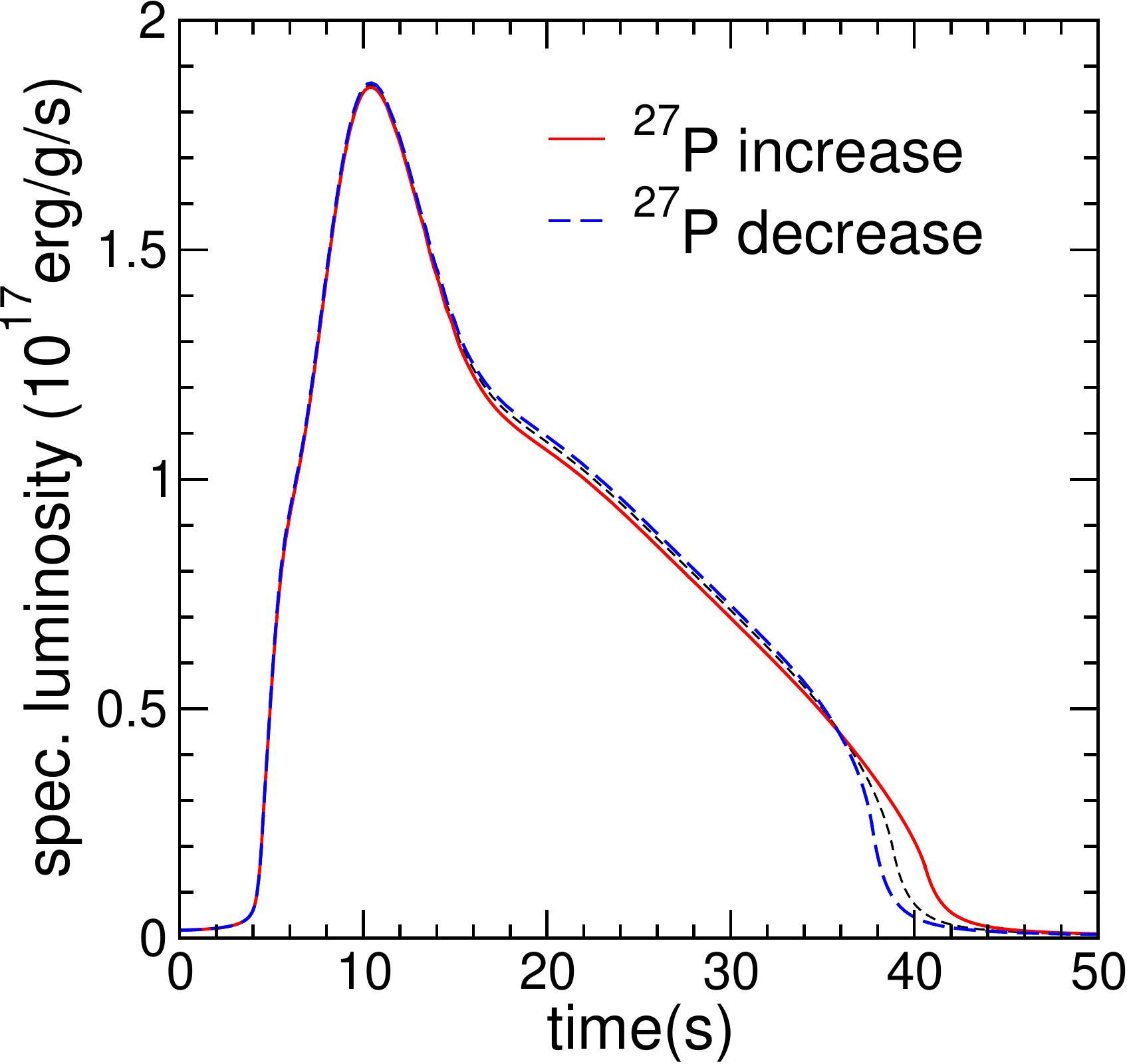}%
\hfill
\includegraphics*[clip=true,width=5.3cm]{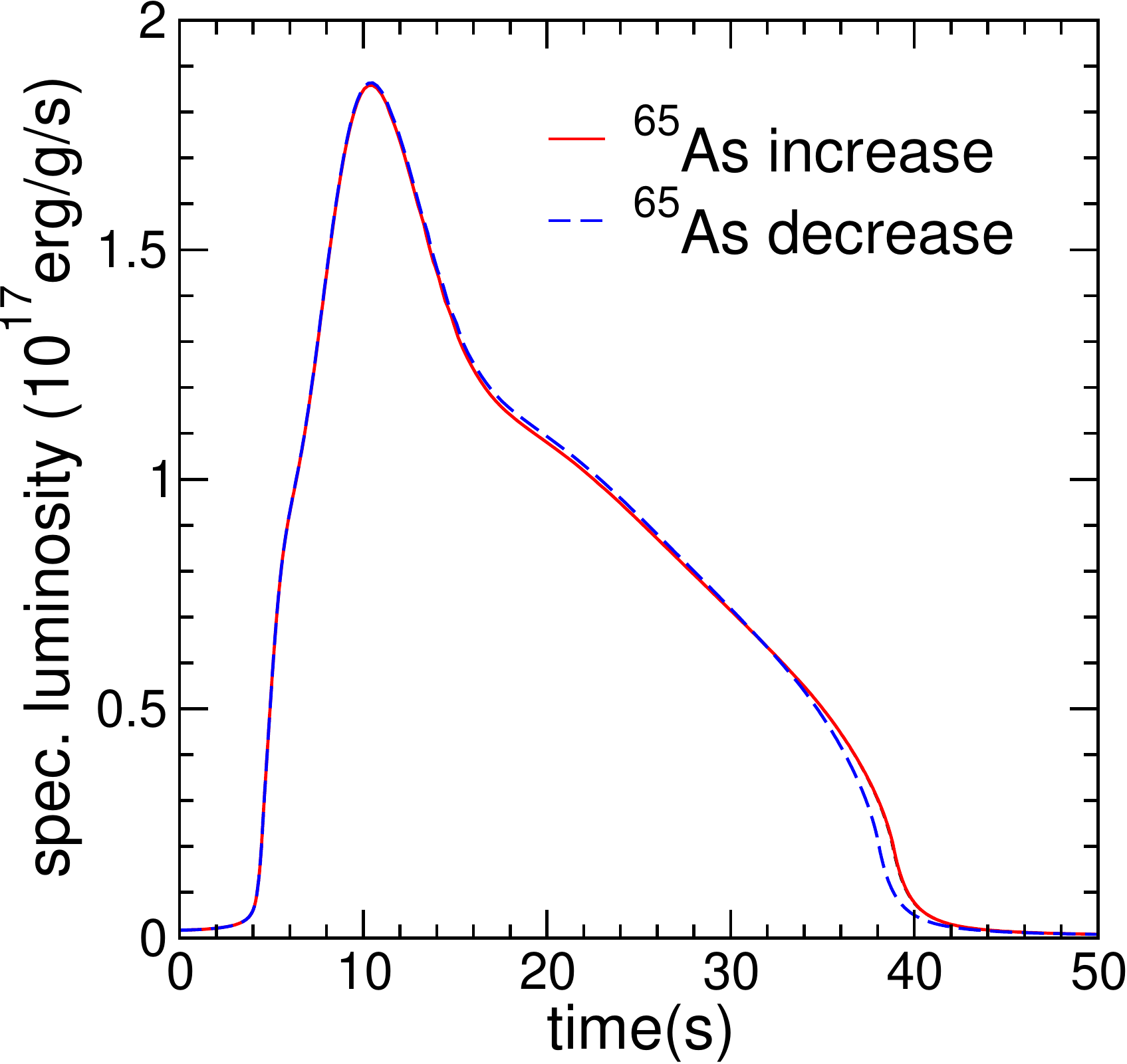}
\end{center}
\caption{\label{FigLC_B} Model B specific luminosities as functions of time for the most significant light curve variations, see Fig.~\ref{FigLC_A} for details.}
\end{figure*}

\begin{figure}
\plotone{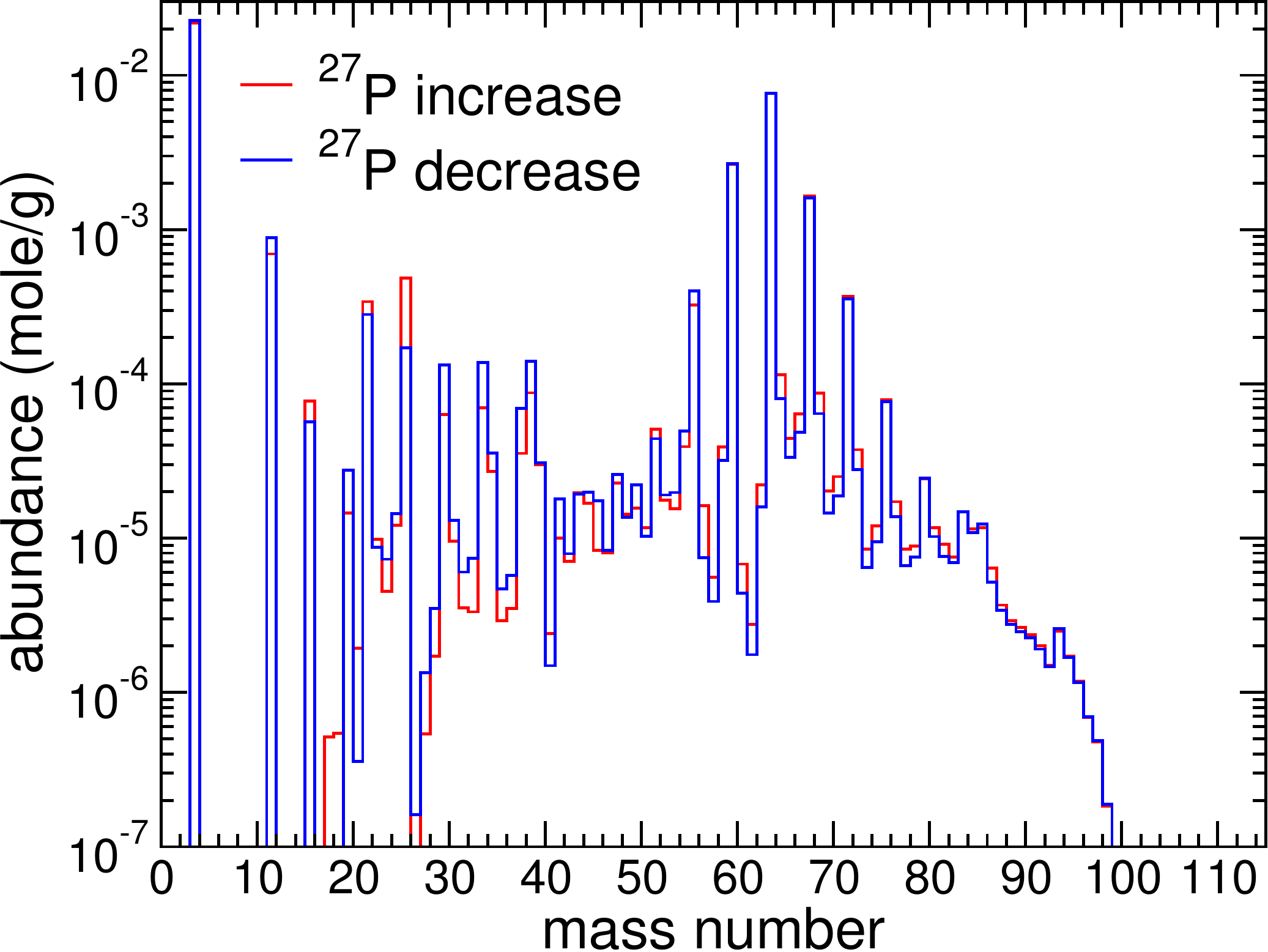}
\caption{\label{FigAB_B} Model B final compositions of the burst ashes summed by mass number for the variation of the $^{27}$P mass.}
\end{figure}

Mass uncertainties affecting Model B, which is characterized by a moderate rp-process, are listed in Tab.~\ref{tbl:LC} and Tab.~\ref{tbl:AB_B}. Only the mass uncertainties of $^{61}$Ga and $^{27}$P have a strong (category 1)  effect on the light curve (Fig.~\ref{FigLC_B}). $^{65}$As has a smaller impact (Fig.~\ref{FigLC_B}). The AME2012 uncertainty of the $^{31}$Cl mass of 50 keV also has a small effect on the light curve ($r_{\rm LC}=2.0$), however, the $^{31}$Cl mass has recently been measured with 3.4~keV precision eliminating this uncertainty \citep{Kankainen2016}. 10 mass uncertainties affect the composition of the burst ashes by more than 20\% (Tab.~\ref{tbl:AB_B}) and an example is shown in Fig.~\ref{FigAB_B}.

\section{Discussion of Light Curves}
Owing to the large number of precision mass measurements of very neutron deficient isotopes in the past decade, the number of remaining nuclear mass uncertainties that affect X-ray burst light curve predictions are now relatively small (see Tab.~\ref{tbl:LC}). For the typical mixed H/He burst of model B, the 38~keV uncertainty of $^{61}$Ga affects the light curve the strongest (when varied by 3$\sigma$). This uncertainty may be underestimated. \citet{Tu2011b} measured the $^{61}$Ga mass using the storage ring technique with an error of 55 keV. However, $^{61}$Ga is at the upper end of the mass range covered, and systematic errors may be significant. The 38 keV uncertainty in AME2012 is derived from combining this result with the measurement of the $^{61}$Ga electron capture Q-value by \citet{Weissman2002}, who quote an uncertainty of 50 keV. However, errors quoted for $\beta$-endpoint measurements have often been demonstrated to be unreliable (for example \citet{Hager2006}). Given the importance of $^{61}$Ga, the insufficient accuracy of the currently recommended mass value, and the potential of underestimated systematic errors, a new measurement of the $^{61}$Ga mass with keV accuracy would be particularly important. 

Only two additional mass uncertainties, $^{27}$P and $^{65}$As, contribute to the light curve uncertainty in model B. $^{27}$P governs the ratio of proton capture on $^{26}$Si to the inverse ($\gamma$,p) photodisintegration on thus produced $^{27}$P and therefore the effective proton capture branch on $^{26}$Si, one of several $\alpha$p-process branch points where proton capture competes with ($\alpha$,p) reactions \citep{Schatz1999a,Fisker2008,Cyburt2016}. $^{65}$As plays the same role for $^{64}$Ge, one of the major waiting points in the rp-process where proton capture competes with $\beta$-decay.  

For model A, which represents an extremely hydrogen rich burst, the ensuing rp-process extends beyond $^{64}$Ge up to the Te region where masses are less well known. Consequently it is affected by more mass uncertainties. The most important mass uncertainties affecting the light curve are 
$^{65}$As and $^{66}$Se. $^{65}$As affects the proton capture branch on the $^{64}$Ge waiting point, similar to model B. However, at the higher temperatures reached in model A, (p,$\gamma$)-($\gamma$,p) equilibrium is not only established between $^{64}$Ge and $^{65}$As, but also with $^{66}$Se. Therefore, the $^{66}$Se mass also affects the effective $^{64}$Ge lifetime in the rp-process \citep{Schatz1998}. 

Another mass uncertainty that strongly affects model A is $^{80}$Zr. This is in part due to the large uncertainty of 1.49 MeV adopted for this isotope in AME2012, which is an artifact of including a low accuracy experiment \citep{Lalleman2001} into the compilation. At this level of uncertainty, theoretical models that have not been considered in the recommended mass, should in principle become competitive. However, theoretical errors are difficult to quantify as $^{80}$Zr lies in a region of strong deformation, and mass predictions for $N=Z$ nuclei have the added complication of the need for a Wigner term that results in enhanced binding (see for example \citet{Goriely2010}). For all these reasons, a measurement of the $^{80}$Zr mass with much improved accuracy would be important. 

Two other important uncertainties for the light curve in model A are $^{58}$Zn and $^{62}$Ge. These nuclei play the same role for the  $^{56}$Ni and $^{60}$Zn waiting points as  $^{66}$Se for the $^{64}$Ge waiting point (see discussion above).  While $r_{\rm LC}$ is not particularly large, these mass uncertainties significantly affect the shape of the early cooling part of the burst light curve. Addressing these uncertainties would therefore be important for attempts to use the shape of this part of the light curve to constrain neutron star properties \citep{Zamfir2012}.

\section{Discussion of Composition}

A larger number of mass uncertainties affect the calculation of the composition of the burst ashes (Tabs.~\ref{tbl:AB_A} and \ref{tbl:AB_B}). The synthesis of $A=79$ nuclei is particularly strongly affected. In model A, the $A=79$ abundance varies by three orders of magnitude from very small (0.01\% mass fraction), to being one of the most important components in the composition (8\% mass fraction, compare to the largest mass fraction of 23\% for $A=105$). In model B, the $A=79$ variation is smaller but still among the largest sensitivities, almost a factor of 10 (0.06\% to 0.6\% mass fraction). The strong sensitivity of the $A=79$ abundance to nuclear masses is primarily due to the large uncertainty of the $^{80}$Zr mass discussed above, but also due to $^{79}$Y (at least in model A). This is an important problem as the amount of odd mass nuclei in the burst ashes directly determines the amount of nuclear Urca-cooling in the outer neutron star crust \citep{Schatz2014}. There are a number of other important odd mass chains that are affected by mass uncertainties. The most abundant ones in model B are $A=65,67,69$. These suffer from 30-40\% uncertainties due to the mass uncertainty of $^{27}$P and, to a lesser extent, $^{61}$Ga. The most abundant odd mass chains in model A are $A=103$ and 105. While $A=105$ is not affected by mass uncertainties, $A=103$ suffers a significant factor of 2 uncertainty due to the mass of $^{65}$As.  The overall most strongly affected mass numbers, besides $A=79$, are $A=82$ and $A=90$ in model A, which are uncertain by an order of magnitude due to mass uncertainties in $^{83}$Nb and $^{91}$Rh, respectively. 

At first sight, the list of important mass uncertainties in this paper differs significantly from previous work \citep{Parikh2009} despite the use of similar thermodynamic trajectories. In fact \citet{Parikh2009} used our model A trajectory among others. Only 5 out of 15 uncertain Q-values  listed in Table IV of \citet{Parikh2009} are affected by mass uncertainties identified in this work. On the other hand, this work identifies 22 additional mass uncertainties that affect composition by more than a factor of 2 - the same criterion used in \citet{Parikh2009}. There are two chief reasons for these differences. First, many new mass measurements have been carried out since AME2003, the mass table used in \citet{Parikh2009}, and therefore many mass uncertainties have been eliminated. Second, \citet{Parikh2009} limited their study to Q-values with $Q<1$~MeV. However, we find that there are many additional cases where photodisintegration rates for Q-values in the 1-2~MeV range are significant. Examples where mass sensitivities identified in this work affect $Q_{(p,\gamma)}$ values above 1~MeV include the $^{65}$As(p,$\gamma$)$^{66}$Se (2.18~MeV), $^{79}$Y(p,$\gamma$)$^{80}$Zr (1.56~MeV),  $^{81}$Zr(p,$\gamma$)$^{82}$Nb (1.09~MeV), $^{82}$Zr(p,$\gamma$)$^{83}$Nb (1.76~MeV),  $^{90}$Ru(p,$\gamma$)$^{91}$Rh (1.14~MeV),  $^{99}$Cd(p,$\gamma$)$^{100}$In (1.67~MeV), and $^{100}$Cd(p,$\gamma$)$^{101}$In (1.71~MeV) reactions ($Q_{(p,\gamma)}$ values are given in parenthesis).  

There are only a few cases where differences remain unexplained by these arguments. We find a strong sensitivity to the $^{27}$P mass uncertainty related to the $\alpha$p-process waiting point $^{26}$Si that is not found in \citet{Parikh2009}. The most likely explanation is that the impact of $\alpha$p-process waiting points tends to depend strongly on the detailed temperature profile. The timing of the temperature rise defines the narrow time window where its hot enough for the reaction flow to have reached the waiting point, and for ($\gamma$,p) reactions to impede further proton capture, but where its still cold enough for the ($\alpha$,p) reaction to be slow. In particular for a rapid temperature rise this time window may disappear altogether. This may also explain, why we only find this sensitivity in our model B and not in model A. While \citet{Parikh2009} used our model A trajectory, they did not use our exact model B trajectory. Another difference to \citet{Parikh2009} is that we do not find a sensitivity to the $^{50}$Fe(p,$\gamma$) Q-value. However, \citet{Parikh2009} find this sensitivity only in one of their trajectories that was artificially scaled to simulate a shorter burst. Lastly, unlike \citet{Parikh2009}, it is found here that $^{97}$Cd and $^{98}$In are important even though the $Q$-value for $^{97}$Cd(p,$\gamma$) is only 0.58 MeV and should therefore be included in  their study. It is not clear what the reason for this discrepancy is. 

\section{Nuclear Physics Uncertainties}

In this work we use X-ray burst models that employ a mass table using calculated Coulomb shifts to predict masses beyond the $N=Z$ line with an accuracy of the order of 100~keV, in addition to the mass uncertainty of the neutron rich mirror nucleus.  It is in principle possible to use the isobaric mass multiplet equation (IMME) to further reduce these uncertainties. The IMME relates the energy states of nuclei within an isospin multiplet, the isobaric analogue states, and has been shown, in cases of mass numbers up to 44, to be able to predict masses to within several tens of keV precision \citep{MacCormick2014}, better than the 100 keV typical of Coulomb shift calculations. In cases where not all members of an isospin multiplet are experimentally determined, unknown IMME coefficients can be calculated with a semi-global fit function utilizing the homogeneous charged sphere approximation. The unknown mass can then be determined from known masses and excitation energies of other members of the multiplet.  For most of the masses of interest in this paper, insufficient experimental information on the isobaric analogue states prevents the IMME from being a useful tool. 

\begin{figure}
\plotone{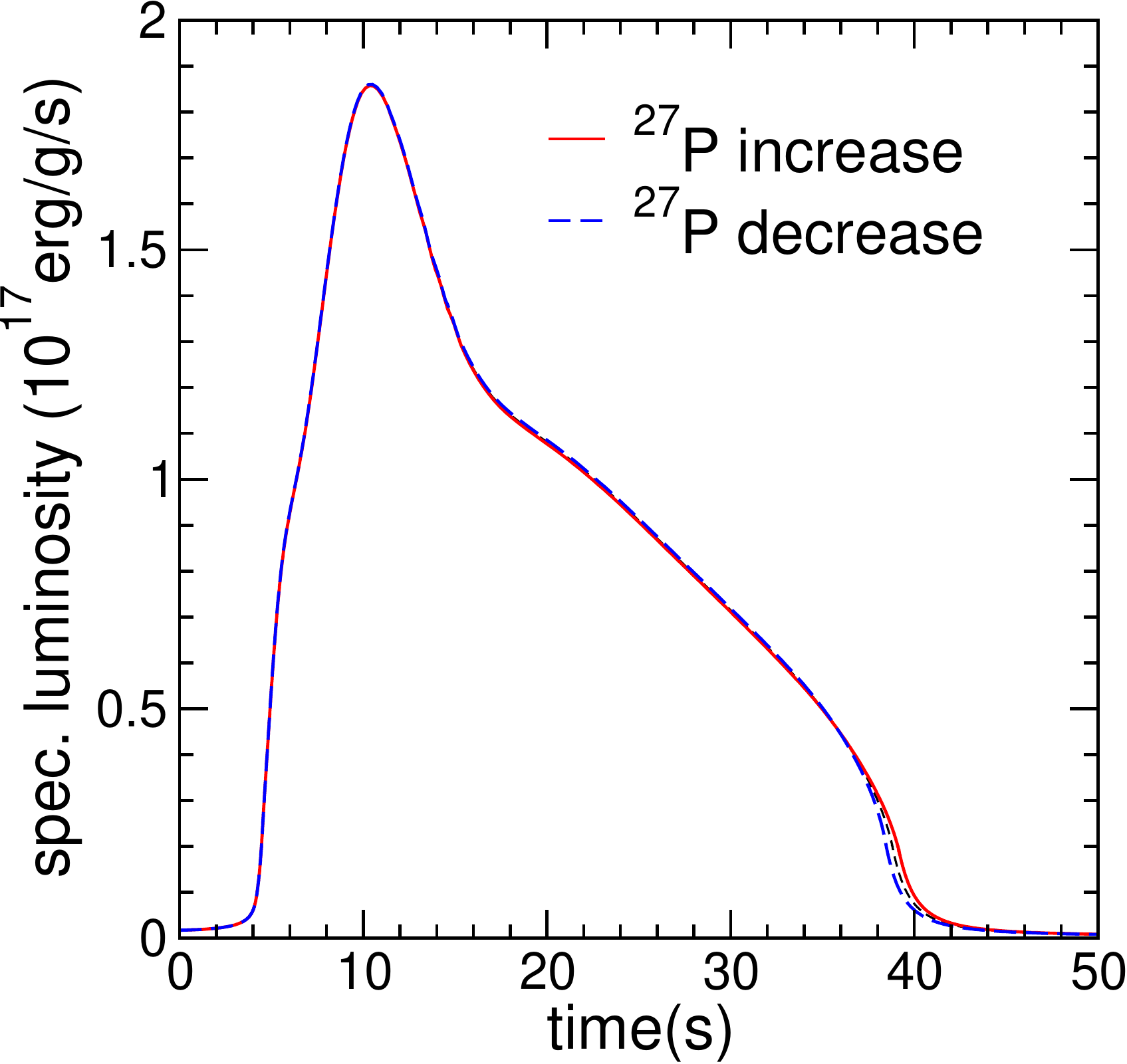}
\caption{\label{FigLC_P27} Model A light curves for the 3$\sigma$ variation of the $^{27}$P mass with reduced $\sigma=$26~keV.}
\end{figure}

The exceptions are the two lightest nuclei, $^{27}$P and $^{56}$Cu. The $^{27}$P ground state is part of the $A~=~27$, $T~=~3/2$ isospin quadruplet, where the masses of the other three members, the ground state of $^{27}$Mg and the analogue states in $^{27}$Al and $^{27}$Si have been precisely measured with uncertainties of less than a few keV. We fit the IMME to these three masses and calculated the mass excess of $^{27}$P to be -716(7) keV. The uncertainty was obtained by varying the three masses within their uncertainties using a Monte Carlo approach and represents a significant improvement over the the AME2012 value of 26~keV. We recommend therefore to use this value in X-ray burst calculations. Test calculations using the smaller uncertainty show, as expected, a much reduced sensitivity to the $^{27}$P mass uncertainty (Fig.~\ref{FigLC_P27}). Nevertheless, even the much smaller uncertainty still leads to an up to 30\% uncertainty in some of the final abundances. There is also the possibility of isospin mixing affecting the validity of the IMME \citep{Wrede2009a}. A precision measurement of the $^{27}$P mass with an accuracy of the order of 1~keV would therefore still be helpful. The other case where the IMME can be applied to reduce mass uncertainties is $^{56}$Cu. The  $^{56}$Cu ground state mass, which is part of the $A~=~56$, $T~=~1$ isospin triplet, has been recently calculated
using the IMME to be -38.685(82) MeV by \citet{Ong2016}. Independently, using $\beta$-delayed proton spectroscopy from the $T~=~2$, $J^{\pi}~=~0^+$ state, \citet{Tu2016} have calculated a mass of  -38.697(88) MeV. These error bars are somewhat smaller than the 100~keV error using Coulomb shift calculations, reducing the $A=55$ abundance uncertainty induced by the $^{56}$Cu mass slightly (see Tab.~\ref{tbl:AB_B}). 

\begin{figure}
\plotone{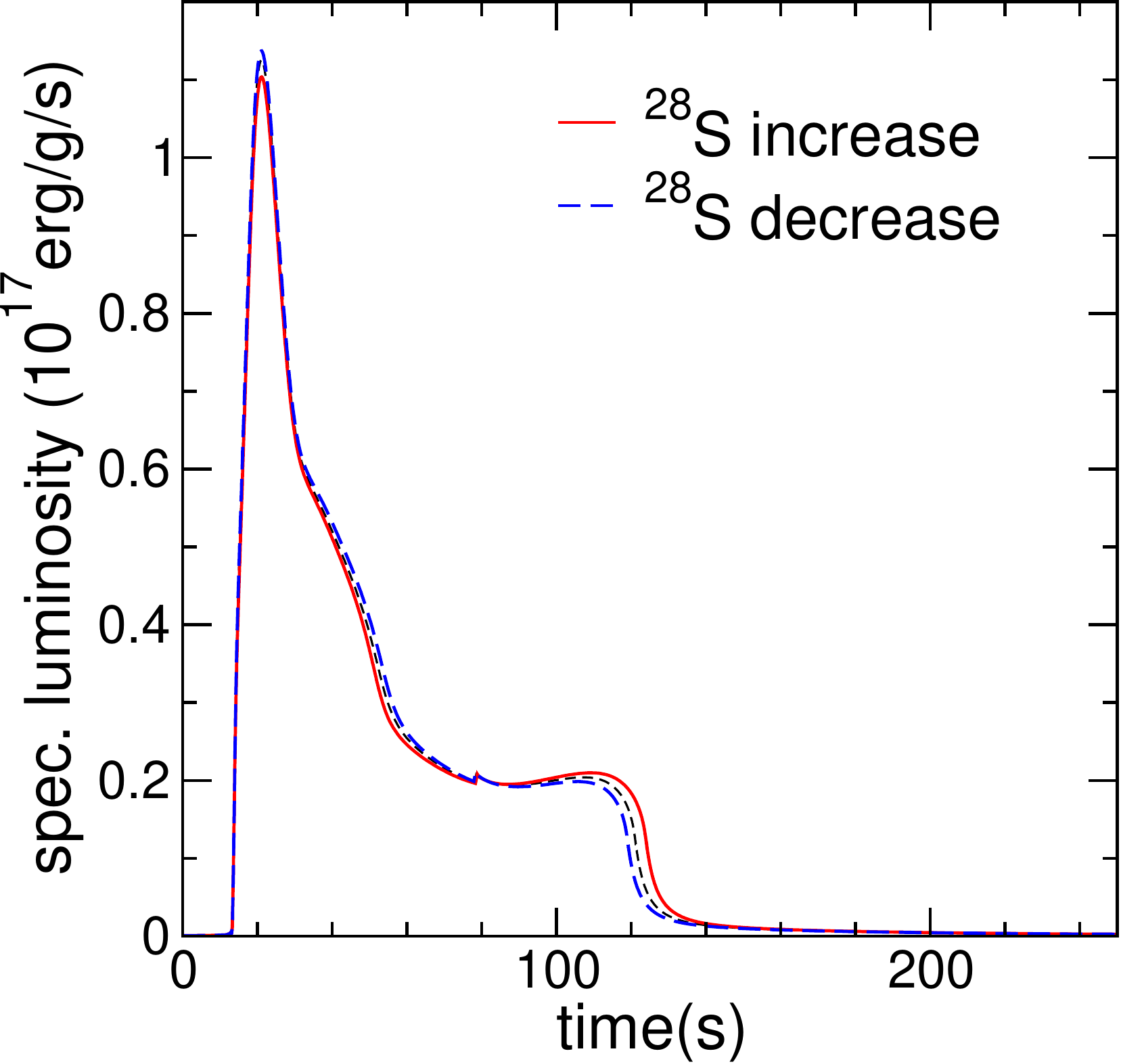}
\caption{\label{FigLC_aplow} Model A light curves for the variation of the $^{28}$S mass when all ($\alpha$,p) rates are reduced by a factor of 100.}
\end{figure}
\begin{figure}
\plotone{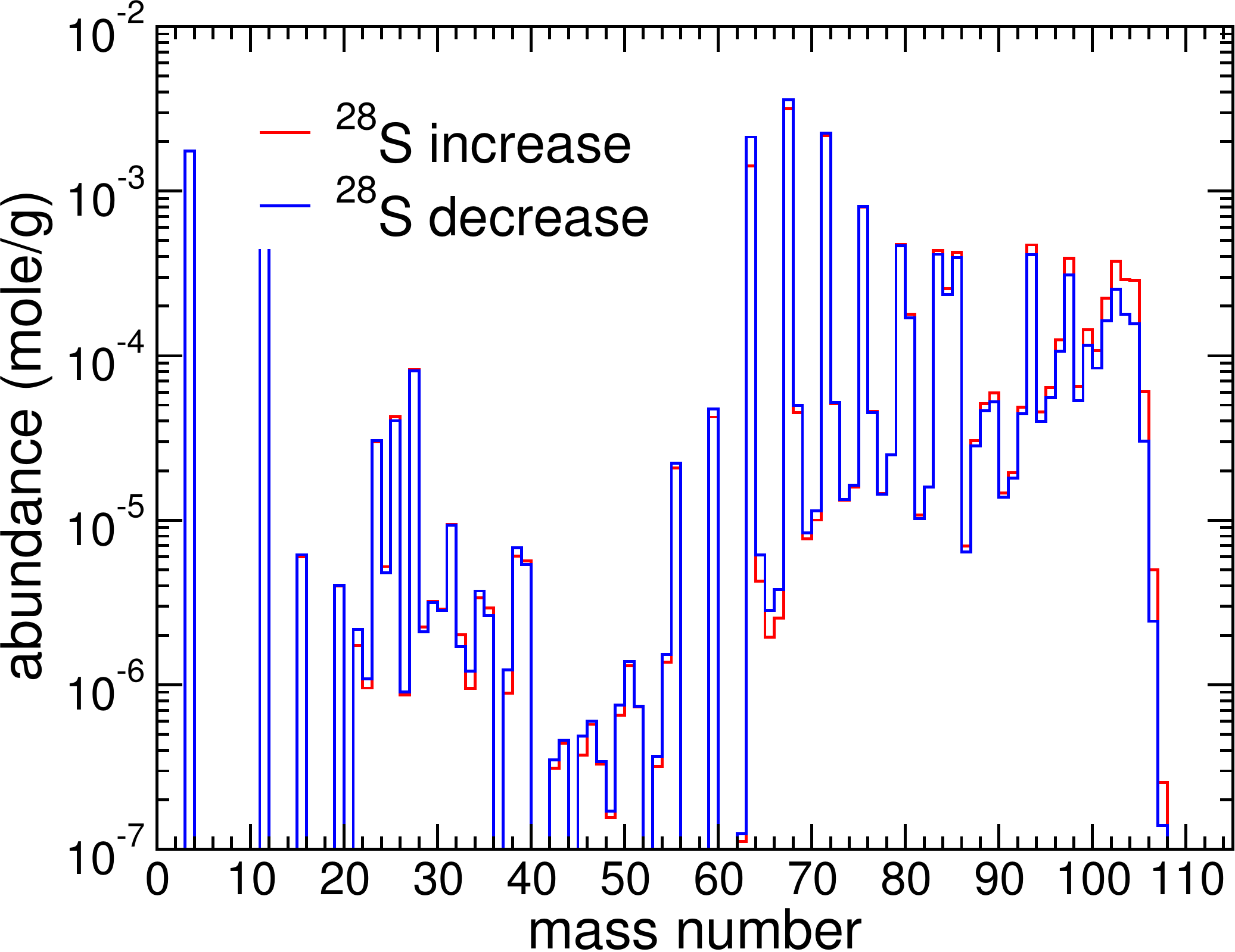}
\caption{\label{FigAB_aplow} Model A final compositions of the burst ashes summed by mass number for the variation of the $^{28}$S mass when all ($\alpha$,p) rates are reduced by a factor of 100.}
\end{figure}

Another significant nuclear uncertainty in x-ray bursts are reaction rates \citep{Parikh2008,Cyburt2016}. A particularly extreme case are ($\alpha$,p) reactions, where recent experimental data on the level structure of the respective compound nuclei indicate that these reaction rates may be systematically overestimated by as much as 1-2 orders of magnitude \citep{Long2016}.  One may ask then the question how robust our results are against future changes of reaction rates due to improved experimental or theoretical data. To explore this question we use this extreme case and repeat the mass sensitivity study for model A with all ($\alpha$,p) reaction rates above Ne, and their inverse, reduced by a factor of 100. Even though the shape of the light curve is changed significantly, the resulting set of masses that affect the light curve remains essentially the same. The main difference is the 160~keV uncertainty of $^{28}$S, which before produced a negligible effect, and has now a significant (Category 1) impact on the light curve (Fig.~\ref{FigLC_aplow}) and composition (Fig.~\ref{FigAB_aplow}). However, there are some differences in the masses affecting the composition of the burst ashes. As expected from the light curve analysis, $^{28}$S now also affects the composition. In addition, there are a number of additional mass uncertainties that would affect the composition in the case of a systematic reduction of ($\alpha$,p) reaction rates, that have not been identified as important using the nominal reaction rates. These are the mass uncertainties of $^{71}$Kr, $^{75}$Sr, $^{84}$Mo, $^{88}$Tc, $^{87}$Ru, $^{93}$Pd, $^{94}$Ag, and $^{96}$Ag. Therefore, while the mass uncertainties affecting the burst light curve appear to be rather robust, in the future some iterative procedure will be needed to identify remaining mass uncertainties that affect the composition of the burst ashes as new reaction rate information becomes available, especially if the changes are large and systematic. Similarly,  reaction rate sensitivity studies may have to be repeated as new mass data become available. 

\section{Conclusions}
This work provides a systematic investigation of the impact of nuclear mass uncertainties on X-ray burst models. Unlike previous studies it uses a self-consistent burst model, instead of a simplified post-processing approach. This enables the investigation of the impact of mass uncertainties on burst light curve predictions, in addition to predictions of the composition of the burst ashes. 
The number of remaining mass uncertainties that need to be addressed is rather small. In a typical mixed H/He burst (Model B) only 3 mass uncertainties have a significant (category 1) effect on the burst light curve ($^{27}$P, $^{61}$Ga, $^{65}$As). Only three additional masses 
($^{80}$Zr, $^{81}$Zr, $^{82}$Nb) affect the composition by more than a factor of 2, and this impact is limited to the $A=79-81$ mass range in the tail end of the composition distribution. In an extreme burst with a maximally extended rp-process (Model A), only 8 masses affect the light curve (category 1 or 2), with $^{65}$As, $^{66}$Se, $^{80}$Zr, $^{91}$Rh, $^{62}$Ge, and $^{58}$Zn being the most relevant (category 1). 11 additional masses along the rp-process in the $A=78-101$ mass range affect the composition by more than a factor of 2. 

The models used in this study span the range from a moderate rp-rpocess reaching into the  $A=60-64$ mass range in model B to the most extreme rp-process in model A. This work therefore likely covers the critical rp-process mass uncertainties for a broad range of models of typical mixed H/He X-ray bursts. This is also supported by the reasonable agreement with the results from \citet{Parikh2009}, who used a broad range of thermodynamic burst model trajectories. Confirming our results with a full 1D X-ray burst model would be useful. In addition, one could expand the investigation to bursts that have less initial hydrogen, but still enough hydrogen to not be dominated by helium burning where the additional mass uncertainty from competition of forward to reverse reaction rates investigated here is not expected to occur. 

Additional mass measurements not listed here may be needed as input into some theoretical reaction rate calculations until direct rate measurements become possible. These need to be identified based on sensitivities of burst models to nuclear reaction rates (see \citep{Cyburt2016}) and the mass sensitivity of the particular theoretical approach chosen to predict the critical reaction rates. 

The mass uncertainties identified in this work as significant (at the 3$\sigma$ level) range from 16~keV (for $^{85}$Mo) to 1.49~MeV (Tab.~\ref{tbl:LC}, \ref{tbl:AB_A}, \ref{tbl:AB_B}). This indicates that a mass accuracy of much better than 10 keV should in most cases be sufficient to ensure mass uncertainties do not contribute significantly (at the 3$\sigma$ level)  to X-ray burst model errors. 
In the near future we can expect a significant enhancement in experimental capabilities to measure masses of extremely neutron deficient nuclei. The new MR-TOF  \citep{Schury2014} and RI-Ring \citep{Yamaguchi2013} devices will enable precision mass measurments at the RIKEN/RIBF rare isotope beam facility, and Penning traps for precision mass measurements at the next generation rare isotope beam facilities FRIB \citep{Redshaw2013} and FAIR \citep{Rodriguez2010} should become online in the next 5-10 years. With these capabilities, all nuclear masses identified in this study should easily be within reach for a sufficiently accurate measurement. This work provides a roadmap for eliminating mass uncertainties in X-ray burst models in light of these developments. 

\section{Acknowledgments}
This material is
based upon work supported by the ISSI International Space Science Institute in Bern, Switzerland and the 
US National Science Foundation under
Grant Numbers PHY-08-22648, and PHY-1430152 (JINA
Center for the Evolution of the Elements).  
The authors thank K. Schmidt for pointing out the potential dependence of the results on changes in ($\alpha$,p) reaction rates, F.-K.\ Thielemann for providing the network solver, and
L.\ Bildsten for contributions to the one-zone model. 

\bibliographystyle{apj}
\bibliography{hsref_v4}

\clearpage
\LongTables
\begin{deluxetable*}{crlrc}
\tablecaption{\label{tbl:AB_A} Impact of mass uncertainties on composition in model A \tablenotemark{a}
}  
\tablewidth{0pt} 
\tablehead{  \colhead{Isotope} & \colhead{$\sigma$\tablenotemark{b}} & Source\tablenotemark{c} & $A$\tablenotemark{d} & $r_{\rm Comp}$
}
\startdata
\iso{Ga}{ 61} &     38 & Exp &  60 &    2.3 \\
\iso{As}{ 65} &     85 & Exp &  26 &    1.7 \\
              &        &     &  28 &    0.51 \\
              &        &     &  64 &    6.5 \\
              &        &     &  69 &    1.4 \\
              &        &     &  72 &    0.81 \\
              &        &     &  76 &    0.76 \\
              &        &     &  80 &    0.70 \\
              &        &     &  81 &    0.70 \\
              &        &     &  82 &    1.3 \\
              &        &     &  84 &    0.70 \\
              &        &     &  85 &    0.73 \\
              &        &     &  86 &    0.81 \\
              &        &     &  94 &    0.64 \\
              &        &     &  97 &    0.76 \\
              &        &     &  98 &    0.55 \\
              &        &     & 101 &    0.79 \\
              &        &     & 102 &    0.66 \\
              &        &     & 103 &    0.65 \\
              &        &     & 104 &    0.70 \\
              &        &     & 106 &    1.5 \\
              &        &     & 107 &    1.8 \\
              &        &     & 108 &    3.0 \\
              &        &     & 109 &    3.9 \\
              &        &     & 110 &    4.7 \\
\iso{Se}{ 66} &    100 & CDE &  26 &    1.2 \\
              &        &     &  28 &    0.73 \\
              &        &     &  64 &    3.0 \\
              &        &     &  69 &    1.3 \\
              &        &     &  98 &    0.81 \\
              &        &     & 107 &    1.2 \\
              &        &     & 108 &    1.5 \\
              &        &     & 109 &    1.7 \\
\iso{Y }{ 78} &    401 & XTP &  26 &    1.3 \\
              &        &     &  77 &    6.2 \\
              &        &     &  78 &    0.42 \\
              &        &     &  79 &    0.54 \\
              &        &     &  80 &    0.60 \\
              &        &     &  81 &    0.64 \\
              &        &     &  84 &    0.83 \\
\iso{Y }{ 79} &    450 & Exp &  78 &    5.0 \\
              &        &     &  79 &    0.048 \\
\iso{Zr}{ 79} &    461 & CDE &  78 &    2.5 \\
\iso{Zr}{ 80} &   1490 & Exp &  26 &    1.2 \\
              &        &     &  28 &    0.79 \\
              &        &     &  64 &    0.81 \\
              &        &     &  68 &    0.75 \\
              &        &     &  72 &    0.77 \\
              &        &     &  76 &    0.78 \\
              &        &     &  79 &   50. \\
              &        &     &  80 &    0.16 \\
              &        &     &  82 &    1.4 \\
              &        &     &  83 &    1.3 \\
              &        &     &  87 &    1.2 \\
              &        &     & 108 &    1.2 \\
\iso{Zr}{ 81} &    165 & Exp &  81 &    0.16 \\
              &        &     &  82 &    0.56 \\
              &        &     &  84 &    1.4 \\
              &        &     &  85 &    1.3 \\
              &        &     &  86 &    1.2 \\
\iso{Zr}{ 82} &    200 & XTP &  82 &    0.34 \\
              &        &     &  83 &    2.4 \\
\iso{Nb}{ 82} &    298 & XTP &  81 &    8.4 \\
              &        &     &  82 &    1.6 \\
              &        &     &  83 &    1.4 \\
              &        &     & 108 &    1.2 \\
\iso{Nb}{ 83} &    300 & Exp &  82 &   10. \\
              &        &     &  83 &    0.22 \\
              &        &     &  84 &    0.61 \\
              &        &     &  85 &    0.77 \\
\iso{Nb}{ 84} &    300 & XTP &  83 &    1.5 \\
\iso{Mo}{ 85} &     16 & Exp &  85 &    0.82 \\
\iso{Tc}{ 86} &    298 & XTP &  85 &    2.8 \\
\iso{Ru}{ 89} &    298 & XTP &  89 &    0.57 \\
              &        &     &  90 &    0.67 \\
              &        &     &  91 &    0.72 \\
              &        &     &  92 &    0.76 \\
              &        &     &  93 &    0.79 \\
\iso{Rh}{ 90} &    401 & XTP &  89 &    1.8 \\
              &        &     &  90 &    1.5 \\
              &        &     &  91 &    1.4 \\
              &        &     &  92 &    1.3 \\
              &        &     &  93 &    1.2 \\
\iso{Rh}{ 91} &    401 & XTP &  90 &   30. \\
              &        &     &  91 &    3.9 \\
              &        &     &  92 &    2.6 \\
              &        &     &  93 &    2.0 \\
              &        &     &  94 &    1.3 \\
              &        &     &  95 &    1.4 \\
              &        &     &  96 &    1.3 \\
              &        &     &  97 &    1.2 \\
\iso{Ag}{ 95} &    401 & XTP &  94 &   14. \\
              &        &     &  95 &    1.7 \\
              &        &     &  96 &    1.5 \\
              &        &     &  97 &    1.4 \\
\iso{Cd}{ 97} &    298 & XTP &  97 &    0.64 \\
              &        &     &  98 &    0.83 \\
\iso{Cd}{ 98} &     52 & Exp &  98 &    0.44 \\
              &        &     &  99 &    1.6 \\
\iso{In}{ 98} &    196 & XTP &  97 &    1.5 \\
\iso{In}{ 99} &    196 & XTP &  98 &    3.9 \\
              &        &     &  99 &    0.55 \\
\iso{In}{100} &    183 & Exp &  99 &    2.3 \\
              &        &     & 100 &    0.80 \\
              &        &     & 101 &    0.73 \\
\iso{In}{101} &    298 & XTP & 100 &    2.3 \\
              &        &     & 101 &    0.69 \\
              &        &     & 102 &    0.70 \\
\enddata
\tablenotetext{a}{Listed are impacts larger than 20\%} 
\tablenotetext{b}{Mass uncertainty in keV. Masses were varied by 3$\sigma$.} 
\tablenotetext{c}{see Tab.~\ref{tbl:LC}}
\tablenotetext{d}{Affected mass number.}
\end{deluxetable*}

\clearpage
\LongTables
\begin{deluxetable*}{crlrc}
\tablecaption{\label{tbl:AB_B} Impact of mass uncertainties on composition in model B \tablenotemark{a} 
}  
\tablewidth{0pt} 
\tablehead{  \colhead{Isotope} & \colhead{$\sigma$\tablenotemark{b}} & Source\tablenotemark{c}& $A$\tablenotemark{d} & $r_{\rm Comp}$
}
\startdata
\iso{P }{ 27} &     26\tablenotemark{e}  & Exp &  12 &    0.79 \\
              &        &     &  16 &    1.4 \\
              &        &     &  20 &    0.53 \\
              &        &     &  22 &    1.2 \\
              &        &     &  26 &    2.8 \\
              &        &     &  30 &    0.48 \\
              &        &     &  31 &    0.73 \\
              &        &     &  34 &    0.51 \\
              &        &     &  35 &    0.76 \\
              &        &     &  38 &    0.51 \\
              &        &     &  39 &    0.62 \\
              &        &     &  42 &    0.56 \\
              &        &     &  46 &    0.48 \\
              &        &     &  50 &    0.71 \\
              &        &     &  54 &    0.78 \\
              &        &     &  55 &    0.79 \\
              &        &     &  56 &    0.81 \\
              &        &     &  57 &    2.2 \\
              &        &     &  59 &    1.2 \\
              &        &     &  63 &    1.4 \\
              &        &     &  65 &    1.4 \\
              &        &     &  66 &    1.3 \\
              &        &     &  67 &    1.3 \\
              &        &     &  69 &    1.4 \\
              &        &     &  70 &    1.4 \\
              &        &     &  71 &    1.3 \\
              &        &     &  73 &    1.3 \\
              &        &     &  75 &    1.3 \\
              &        &     &  77 &    1.3 \\
\iso{Cu}{ 56} &    100\tablenotemark{f} & CDE &  55 &    1.6 \\
\iso{Ga}{ 61} &     38 & Exp &  12 &    0.69 \\
              &        &     &  20 &    0.71 \\
              &        &     &  42 &    0.82 \\
              &        &     &  46 &    0.78 \\
              &        &     &  57 &    1.3 \\
              &        &     &  60 &    3.6 \\
              &        &     &  61 &    3.7 \\
              &        &     &  63 &    2.3 \\
              &        &     &  64 &    0.69 \\
              &        &     &  65 &    0.83 \\
              &        &     &  66 &    0.82 \\
              &        &     &  67 &    0.83 \\
              &        &     &  68 &    0.77 \\
\iso{As}{ 65} &     85 & Exp &  68 &    0.75 \\
              &        &     &  69 &    0.78 \\
              &        &     &  70 &    0.78 \\
              &        &     &  71 &    0.78 \\ 
              &        &     &  72 &    0.67 \\
              &        &     &  73 &    0.70 \\
              &        &     &  75 &    0.70 \\
              &        &     &  76 &    0.68 \\
              &        &     &  77 &    0.71 \\
              &        &     &  79 &    0.71 \\
              &        &     &  80 &    0.71 \\
              &        &     &  81 &    0.75 \\
              &        &     &  82 &    0.77 \\
              &        &     &  84 &    0.79 \\
\iso{Y }{ 78} &    401 & XTP &  77 &    1.8 \\
\iso{Zr}{ 79} &    461 & CDE &  78 &    1.5 \\
\iso{Zr}{ 80} &   1490 & Exp &  79 &    9.3 \\
              &        &     &  80 &    0.17 \\
              &        &     &  81 &    0.56 \\
              &        &     &  84 &    0.53 \\
              &        &     &  85 &    0.52 \\
              &        &     &  86 &    0.48 \\
\iso{Zr}{ 81} &    165 & Exp &  81 &    0.40 \\
              &        &     &  82 &    0.80 \\
              &        &     &  84 &    1.3 \\
              &        &     &  85 &    1.3 \\
              &        &     &  86 &    1.4 \\
\iso{Nb}{ 82} &    298 & XTP &  81 &    3.3 \\
              &        &     &  82 &    1.3 \\
              &        &     &  84 &    0.72 \\
              &        &     &  85 &    0.68 \\
              &        &     &  86 &    0.61 \\
\iso{Nb}{ 83} &    300 & Exp &  82 &    1.5 \\
\iso{Tc}{ 86} &    298 & XTP &  85 &    1.5 \\
\enddata
\tablenotetext{a}{Listed are impacts larger than 20\%} 
\tablenotetext{b}{Mass uncertainty in keV. Masses were varied by 3$\sigma$.} 
\tablenotetext{c}{see Tab.~\ref{tbl:LC}}
\tablenotetext{d}{Mass number affected.}
\tablenotetext{e}{Error using IMME: 7 keV (see text).}
\tablenotetext{f}{Error using IMME: 82 keV \citep{Ong2016}}
\end{deluxetable*}

\end{document}